\documentclass[aps,prx,onecolumn,amsmath,amssymb]{revtex4-1}

\usepackage{graphicx}
\usepackage{dcolumn}
\usepackage{bm}
\usepackage{epsfig}
\usepackage{subfig}
\usepackage[all]{xy}
\usepackage{amsmath}
\usepackage{amsfonts}

\newcommand{\ba}{\begin{eqnarray}}
\newcommand{\ea}{\end{eqnarray}}
\newcommand{\be}{\begin{equation}}
\newcommand{\ee}{\end{equation}}
\newcommand{\bea}{\begin{eqnarray}}
\newcommand{\eea}{\end{eqnarray}}

\usepackage{theorem}

\theorembodyfont{\rmfamily}

\theoremstyle{break}

\def\QED{~\rule[-1pt]{5pt}{5pt}\par\medskip}

\def\Re {\mathbb{R}}

\begin{document}

\title{Fidelity and Fisher information on quantum channels}
\author{Haidong Yuan}
\email{hdyuan@mae.cuhk.edu.hk}
\affiliation{Department of Mechanical and Automation Engineering, The Chinese University of Hong Kong, Shatin, Hong Kong}

\author{Chi-Hang Fred Fung}
\email{fred.fung@huawei.com}
\affiliation{
Munich Research Center, Huawei Technologies D\"usseldorf GmbH, Munich, Germany
}

\date{\today}

\begin{abstract}
The fidelity function of quantum states have been widely used in quantum information science and frequently arises in the quantification of optimal performances for the estimation and distinguish of quantum states. A fidelity function on quantum channel is expected to have same wide applications in quantum information science. In this paper we propose a fidelity function on quantum channels and show that various distance measures on quantum channels can be obtained from this fidelity function, for example the Bures angle and the Bures distance can be extended to quantum channels via this fidelity function. We then show that the distances between quantum channels lead naturally to a new Fisher information which quantifies the ultimate precision limit in quantum metrology, the ultimate precision limit can thus be seen as a manifestation of the distances between quantum channels. We also show that the fidelity on quantum channels provides a unified framework for perfect quantum channel discrimination and quantum metrology, in particular we show that the minimum number of uses needed for perfect channel discrimination is exactly the counterpart of the precision limit in quantum metrology, and various useful lower bounds for the minimum number of uses needed for perfect channel discrimination can be obtained via this connection.
\end{abstract}

\maketitle
             
\section{Introduction}

Fidelity, as a measure of the distinguishability between quantum states\cite{Fuchs1994,Jozsa1994,Fuchs1996}, plays an important role in many areas of quantum information science, for example it is related to the precision limit in quantum metrology~\cite{BRAU94},
serves as a measure of entanglement preservation through noisy quantum channels~\cite{Schumacher1996},
and a measure of entanglement preservation in quantum memory~\cite{Surmacz2006};
it has also been used as a characterization method for quantum phase transitions~\cite{Gu2010},
and a criterion for successful transmission in formulating quantum channel capacities~\cite{Barnum1998}. 

Unlike the fidelity of quantum states which is defined directly on quantum states, most commonly used measures for the distinguishability of quantum channels are defined indirectly through the effects of the channels on the states. For example the diamond norm, which is defined as $\|K_1-K_2\|_{\diamond}=\max_{\rho_{SA}}\|K_1\otimes I_A(\rho_{SA})-K_2\otimes I_A(\rho_{SA})\|_1$\cite{Kitaev1997,Kitaev2002,watrous2009}( here $\|X\|_1=Tr\sqrt{X^\dagger X}$, $\rho_{SA}$ denotes a state on system+ancilla, and $I_A$ denotes the identity operator on the ancillary system), is induced by the trace distance on quantum states $\|\rho_1-\rho_2\|_1$; another measure on quantum channels which is defined as $\arccos F_{\min}(K_1,K_2)=\arccos \min_{\rho_{SA}} F_S[K_1\otimes I_A(\rho_{SA}), K_2\otimes I_A(\rho_{SA})]$\cite{Gilchrist,Belavkin}, is induced by the fidelity on quantum states $F_S(\rho_1,\rho_2)=Tr\sqrt{\rho_1^{\frac{1}{2}}\rho_2\rho_1^{\frac{1}{2}}}$. These induced measures through quantum states lack a direct connection to the properties of quantum channels, which severely restrict the insights that can be gained from these measures. A direct measure on quantum channels is expected to provide more insights thus highly desired.

In this paper we provide a fidelity function defined directly on quantum channels, and show that this fidelity function on quantum channels, together with the classical fidelity on probability distribution and the fidelity on quantum states, form a hierarchy of fidelity functions in terms of optimization. This fidelity function on quantum channels also lead to various distance measures defined directly on quantum channels, in particular we show the Bures angle and the Bures distance can be extended to quantum channels. We then show the distance between quantum channels leads naturally to a new Fisher information on quantum channels which quantifies the ultimate precision limit in quantum metrology. We also show that this fidelity function provides a unified framework for perfect quantum channel discrimination and quantum metrology, in particular we show the minimum number of uses needed for perfect channel discrimination is exactly the counterpart of the precision limit in quantum metrology, and various useful lower bounds for the minimum number of uses needed for perfect channel discrimination can be obtained via this connection.

\section{Fidelity function on quantum channels}

We start by defining the fidelity function on unitary channels then extend it to noisy channels.

For a $m\times m$ unitary matrix $U$, we denote $e^{-i\theta_j}$ as the eigenvalues of $U$, where $\theta_j\in(-\pi,\pi]$ for $1\leq j\leq m$ and we call $\theta_j$ the eigen-angles of $U$. We define(see also\cite{Chau2011,Fung1,Fung2})
$\parallel U\parallel_{\max}=\max_{1\leq j \leq m}\mid\theta_j\mid,$
 and $\parallel U\parallel_g$ as the minimum of $\parallel e^{i\gamma}U\parallel_{\max}$ over equivalent unitary operators with different global phases, i.e., $\parallel U\parallel_g=\min_{\gamma\in \Re} \parallel e^{i\gamma} U\parallel_{\max}$. We then define
\begin{eqnarray}
C(U)=\left\{\begin{array}{cc}
\parallel U\parallel_g, & if \parallel U\parallel_g\leq \frac{\pi}{2}, \\
\frac{\pi}{2}, & if \parallel U\parallel_g> \frac{\pi}{2}.\\
\end{array}\right.
\end{eqnarray}
Quantitatively $C(U)$ is equal to the maximal angle that $U$ can rotate a state away from itself\cite{Acin01,Mauro2001,Fung2}, i.e.,
$\cos[C(U)]=\min_{|\psi\rangle}|\langle \psi |U |\psi \rangle|.$
For mixed states it can be written as
$\cos[C(U)]=\min_{\rho}F_S(\rho, U\rho U^\dagger).$

If $\theta_{\max}=\theta_1\geq \theta_2\geq \cdots \geq \theta_m=\theta_{\min}$ are arranged in decreasing order, then $C(U)=\frac{\theta_{\max}-\theta_{\min}}{2}$ when $\theta_{\max}-\theta_{\min}\leq \pi$\cite{Fung2}.
We then define $\Theta_{QC}(U_1,U_2)=C(U_1^\dagger U_2)$, here $U_1$ and $U_2$ are unitary operators on the same Hilbert space(we can expand the space if they are not the same).
It is easy to see that
\begin{eqnarray}
\aligned
\cos[\Theta_{QC}(U_1,U_2)]&=\cos[C(U_1^\dagger U_2)]\\
 &=\min_{\rho}F_S(U_1\rho U_1^\dagger, U_2\rho U_2^\dagger),
\endaligned
\end{eqnarray}
$\Theta_{QC}(U_1,U_2)$ thus corresponds to the maximal angle between the output states of $U_1$ and $U_2$(however we note that the definition of $\Theta_{QC}(U_1,U_2)$ is independent of the states). We then denote $F_{QC}(U_1,U_2)=\cos[\Theta_{QC}(U_1, U_2)]$ as the fidelity between $U_1$ and $U_2$. For unitary channels this is equivalent to the fidelity function proposed previously in \cite{Acin01}.

We now generalize this to noisy quantum channels. A general quantum channel $K$, which maps from $m_1$- to $m_2$-dimensional Hilbert space, can be represented by Kraus operators, $K(\rho_S)=\sum_{j=1}^q F_j\rho_S F^\dagger_j$ where
$\sum_{j=1}^q F^\dagger_jF_j=I$. Equivalently it can also be written as
$K(\rho_S)=Tr_E(U_{ES}(|0_E\rangle\langle0_E|\otimes \rho_S) U^\dagger_{ES}),$
where $|0_E\rangle$ denotes some standard state of the environment, and $U_{ES}$ is a unitary operator acting on both system and environment, which we call as the unitary extension of $K$.

We define
$\Theta_{QC}(K_1, K_2)=\min_{\{U_{ES1},U_{ES2}\}}\Theta_{QC}(U_{ES1}, U_{ES2})$ and
$F_{QC}(K_1, K_2)=\cos \Theta_{QC}(K_1, K_2),$
 where $U_{ESi}$ are unitary extensions of $K_i$, $i\in\{1,2\}$. 
In Appendix~\ref{app-compute},
 we show that the 
 optimization can be taken by fixing one unitary extension and just optimizing over the other unitary extension, i.e.,
\begin{eqnarray}
\aligned
\Theta_{QC}(K_1, K_2)&=\min_{U_{ES1}}\Theta_{QC}(U_{ES1}, U_{ES2})\\
&=\min_{U_{ES2}}\Theta_{QC}(U_{ES1}, U_{ES2}).
\endaligned
\end{eqnarray}
In terms of $F_{QC}(K_1,K_2)$ it can be written as
\begin{eqnarray}
\aligned
F_{QC}(K_1, K_2)&=\max_{U_{ES1}}F_{QC}(U_{ES1}, U_{ES2})\\
&=\max_{U_{ES2}}F_{QC}(U_{ES1}, U_{ES2}).
\endaligned
\end{eqnarray}
 This can be seen as the counterpart of Uhlmann's purification theorem on quantum states~\cite{Uhlmann1976}(however the proof does not use Uhlmann's purification theorem~\cite{Yuan2017npj}).
In Appendix~\ref{app-metric},
we show that $\Theta_{QC}(K_1, K_2)$ is a metric and can be computed directly from the Kraus operators of $K_1$ and $K_2$ as~\cite{Yuan2017npj}
\begin{equation}
\label{eq:dk}
\Theta_{QC}(K_1, K_2)=\arccos \max_{\|W\|\leq 1}\frac{1}{2}\lambda_{\min}(K_W+K^\dagger_W),
\end{equation}
 here $\lambda_{\min}(K_W+K^\dagger_W)$ denotes the minimum eigenvalue of $K_W+K^\dagger_W$ with $K_W=\sum_{ij}w_{ij}F_{1i}^\dagger F_{2j}$, $F_{1i}$ and $F_{2j}$ denote the Kraus operators of $K_1$ and $K_2$ respectively, $w_{ij}$ denotes the $ij$-th entry of a $q\times q$ matrix $W$ with $\|W\|\leq 1$ where $\|\cdot\|$ is the operator norm which corresponds to the maximum singular value, here $W$ arises from the non-uniqueness of the Kraus representations. Thus
\begin{equation}
\label{eq:F_QC1}
F_{QC}(K_1,K_2)=\max_{\|W\|\leq 1}\frac{1}{2}\lambda_{\min}(K_W+K^\dagger_W).
 \end{equation}
  We emphasize that $F_{QC}$ is defined directly on quantum channels without referring to the states, such direct connection, in contrast to the induced measure, is crucial when applying the fidelity to channel discrimination and quantum metrology as we will show later.  Furthermore the fidelity can be formulated as a semi-definite programming and computed efficiently as
$\max_{\|W\|\leq 1}\frac{1}{2}\lambda_{\min}(K_W+K^\dagger_W)=$
\begin{eqnarray}
\label{eq:sdp}
\aligned
&max \qquad \frac{1}{2}t \\
s.t.\qquad &\left(\begin{array}{cc}
      I & W^\dagger  \\
      W & I \\
          \end{array}\right)\succeq 0,\\
      &    K_W+K^\dagger_W-tI \succeq 0.
          \endaligned
          \end{eqnarray}

Analogous to the Bures distance on quantum states $B_S(\rho_1,\rho_2)=\sqrt{2-2F_S(\rho_1,\rho_2)}$, we can similarly define a Bures distance on quantum channels as $B_{QC}(K_1,K_2)=\sqrt{2-2F_{QC}(K_1,K_2)}$. 
In Appendix~\ref{app-compute},
we prove an intriguing and useful connection between $B_{QC}(K_1,K_2)$ and the minimum distances between the Kraus operators of $K_1$ and $K_2$ as
\begin{eqnarray}
\aligned
\nonumber
B_{QC}^2(K_1,K_2)
=\min_{\{\tilde{F}_{1i}\},\{\tilde{F}_{2i}\}} \| \sum_{i} (\tilde{F}_{1i}-\tilde{F}_{2i})^\dagger (\tilde{F}_{1i}-\tilde{F}_{2i})\| \\
\endaligned
\end{eqnarray}
where $\{\tilde{F}_{1i}\},\{\tilde{F}_{2i}\}$ are the sets of all equivalent Kraus representations of $K_1$ and $K_2$ respectively. This connection is particular useful in studying the scalings of the distance between quantum channels as we will show later.

In which sense we call $F_{QC}(K_1,K_2)$ a fidelity function?  It turns out that 
 $F_{QC}(K_1,K_2)=\min_{\rho_{SA}}F_S[K_1\otimes I_A (\rho_{SA}), K_2\otimes I_A(\rho_{SA})]$.
To see this, it is proved in the supplemental material of Ref.~\cite{Yuan2017npj} that 
\begin{eqnarray}
\min_{\rho_{SA}}F_S[K_1\otimes I_A (\rho_{SA}), K_2\otimes I_A(\rho_{SA})]
=\max_{\|W\|\leq 1 }\frac{1}{2}\lambda_{\min}(K_{W}+K^\dagger_{W}),
\end{eqnarray}
which coincides with Eq.~\eqref{eq:F_QC1}.
From this relationship it is also immediate clear that $F_{QC}(K_1,K_2)$ is stable, i.e., $F_{QC}(K_1\otimes I,K_2\otimes I)=F_{QC}(K_1,K_2)$.
 This result gives an operational meaning to $F_{QC}(K_1,,K_2)$. We emphasize that although we made connections between $F_{QC}(K_1,K_2)$ and the minimum fidelity of the output states, $F_{QC}(K_1,K_2)$ is defined directly on quantum channels and does not depend on the states. The definition and the operational meaning of $F_{QC}(K_1,K_2)$ play distinct roles in applications, the operational meaning provides a physical picture while the direct definition brings insights which enable or ease the proofs and computations, which will be demonstrated in the applications. This is in analogy to how fidelity of quantum states is connected to the classical fidelity $F_S(\rho_1,\rho_2)=\min_{\{E_i\}}F_C(p_1,p_2)$, here $F_C(p_1,p_2)=\sum_{i}\sqrt{p_{1i}}\sqrt{p_{2i}}$ denotes the classical fidelity with $p_{1i}=Tr(\rho_1E_i)$ and $p_{2i}=Tr(\rho_2E_i)$, $\{E_i\}$ denotes a set of Positive Operator Valued Measurements(POVM)\cite{Fuchs1996}, here similarly the fidelity between quantum states has the operational meaning as the minimum classical fidelity, however the fidelity between quantum states is defined directly on quantum states which is independent of the measurements and such direct definition has provided numerous insights which would be hindered with just the classical fidelity. 
 

 It is known that the trace distance and the fidelity between quantum states have the following relationships\cite{Fuchs1999}
\begin{equation}
1-F_S(\rho_1,\rho_2)\leq \frac{1}{2}\|\rho_1-\rho_2\|_1\leq\sqrt{1-F_S^2(\rho_1,\rho_2)},
\end{equation}
from which it is straightforward to get the relationships between the diamond norm and the fidelity of quantum channels. This can be obtained by substituting $\rho_1=K_1\otimes I_A(\rho_{SA})$ and $\rho_2=K_2\otimes I_A(\rho_{SA})$, then optimizing over $\rho_{SA}$
\begin{eqnarray}
\aligned
\max_{\rho_{SA}}1-F_S[K_1\otimes I_A(\rho_{SA}),K_2\otimes I_A(\rho_{SA})]&\leq \max_{\rho_{SA}}\frac{1}{2}\|K_1\otimes I_A(\rho_{SA})-K_2\otimes I_A(\rho_{SA})\|_1\\
&\leq \max_{\rho_{SA}}\sqrt{1-F_S^2[K_1\otimes I_A(\rho_{SA}),K_2\otimes I_A(\rho_{SA})]},
\endaligned
\end{eqnarray}
which gives
\begin{equation}
1-F_{QC}(K_1,K_2)\leq \frac{1}{2}\|K_1-K_2\|_{\diamond}\leq \sqrt{1-F_{QC}^2(K_1,K_2)}.
\end{equation}
Since $F_{QC}(K_1,K_2)$ can be computed directly from the Kraus operators, this also provides a way to bound the diamond norm using the Kraus operators.

In \cite{Raginsky} the Choi matrices of the quantum channels are used to compute the fidelity between the channels, which corresponds to the fidelity between the output states of two quantum channels when the input state is taken as the maximal entangled state. As the maximal entangled state is in general not the optimal input state, the fidelity thus defined does not have operational meaning as the minimum fidelity of the output states, thus can not be related to the ultimate precision limit in quantum metrology etc(instead related to the precision limit when the probe state is taken as the maximally entangled state).

\section{A unified framework for quantum metrology and perfect channel discrimination}

Next we demonstrate the applications in quantum information science, in particular we show how the fidelity provides a unified platform for the ultimate precision in quantum metrology and the minimum number of uses needed for perfect channel discrimination. 

The task of quantum metrology, or quantum parameter estimation in general, is to estimate a parameter $x$ encoded in some channel $K_x$, this can be achieved by preparing a quantum state $\rho_{SA}$ and let it go through the extended channel $K_x\otimes I_A$ with the output state $\rho_x=K_x\otimes I_A(\rho_{SA})$. By performing POVM, $\{E_y\}$, on $\rho_x$ one gets the measurement result $y$ with probability $p(y|x)=Tr(E_y\rho_x)$. According to the Cram\'{e}r-Rao bound\cite{HELS67,HOLE82,CRAM46,Rao}, the standard deviation for any unbiased estimator of $x$ is bounded below by $\delta \hat{x}\geq \frac{1}{\sqrt{nJ_C[p(y|x)]}},$ where $\delta \hat{x}$ is the standard deviation of the estimation of $x$, $J_C[p(y|x)]$ is the classical Fisher information
and $n$ is the number of times that the procedure is repeated.
The classical Fisher information can be further optimized over all POVMs, which gives
\begin{equation}
\label{eq:J}
\delta\hat{x}\geq\frac{1}{\sqrt{n\max_{\{E_y\}}J_C[p(y|x)]}}=\frac{1}{\sqrt{nJ_S(\rho_x)}},
\end{equation}
 where the optimized value $J_S(\rho_x$) is usually called the quantum Fisher information\cite{HELS67, HOLE82,BRAU94,BRAU96}, here for distinguish we will call it the quantum state Fisher information.

We first recall established connections between the fidelity functions and the Fisher information. Given $\rho_x$ and its infinitesimal state $\rho_{x+dx}$, for a given POVM $\{E_y\}$, the classical fidelity between $p(y|x)=Tr(E_y\rho_x)$ and $p(y|x+dx)=Tr(E_y\rho_{x+dx})$ is given by $F_C[p(y|x),p(y|x+dx)]=\sum_{y_i} \sqrt{p(y_i|x)}\sqrt{p(y_i|x+dx)}$ which defines an angle as $\cos \Theta_C[p(y|x),p(y|x+dx)]=F_C[p(y|x),p(y|x+dx)]$. The classical Fisher information is related to the classical fidelity as $\frac{1}{4}J_C[p(y|x)]dx^2=2-2F_C[p(y|x),p(y|x+dx)]$ up to the second order of $dx$\cite{BRAU94},
this can also be written as
\begin{equation}
\label{eq:prob}
J_C[p(y|x)]=\lim_{dx\rightarrow 0}\frac{4\Theta_C^2[p(y|x),p(y|x+dx)]}{dx^2}.
\end{equation}

If we optimize over $\{E_y\}$ the classical fidelity then leads to the fidelity between quantum states as \cite{BRAU94}
\begin{equation}\min_{\{E_y\}} F_C[Tr(E_y\rho_x), Tr(E_y\rho_{x+dx})]=F_S(\rho_x,\rho_{x+dx}),\end{equation} and the classical Fisher information leads to the quantum state Fisher information $J_S(\rho_x)=\max_{\{E_y\}}J_C[p(y|x)]$ and up to the second order of $dx$\cite{BRAU94,BRAU96}
\begin{eqnarray}
\label{eq:qsf}
\aligned
\frac{1}{4}J_S(\rho_x)dx^2=2-2F_S(\rho_x,\rho_{x+dx}).
\endaligned
\end{eqnarray}
If we denote $\cos \Theta_S(\rho_x,\rho_{x+dx})= F_S(\rho_x,\rho_{x+dx})$, then 
 \begin{eqnarray}
\label{eq:BJ}
\aligned
J_S(\rho_x)&=\lim_{dx\rightarrow 0}\frac{8[1-\cos \Theta_S(\rho_x,\rho_{x+dx})]}{dx^2}\\
&=\lim_{dx\rightarrow 0}\frac{4\Theta_S^2(\rho_x,\rho_{x+dx})}{dx^2}.
\endaligned
\end{eqnarray}

The precision can be further improved by optimizing over the probe states, which leads to the ultimate local precision limit of estimating $x$ from $K_x$. Intuitively, this ultimate precision limit should be quantified by the distance between $K_x$ and its infinitesimal neighboring channel $K_{x+dx}$, in a way analogous to how Bures distance of quantum states quantifies the precision limit of estimating $x$ from the state $\rho_x$\cite{BRAU94}. However although much progress has been made on calculating the ultimate precision limit\cite{Fujiwara2008,Escher2011,Tsang2013,Rafal2012,durkin,Knysh2014,Jan2013,Rafal2014,Alipour2014}, such a clear physical picture has still not been established after more than two decades since Braunstein and Caves's seminal paper\cite{BRAU94}, this is mainly due to the lack of proper tools on quantum channels. Here we show that the fidelity between quantum channels can be used to establish such a physical picture, which also leads naturally to a new Fisher information on quantum channel.

Further optimizing over the probe states
\begin{eqnarray}
\aligned
\max_{\rho_{SA}} \frac{1}{4}J_S(\rho_x)dx^2&=2-2\min_{\rho_{SA}}F_S(\rho_x,\rho_{x+dx})\\
&=2-2F_{QC}(K_x, K_{x+dx})\\
&=B_{QC}^2(K_x,K_{x+dx}),
\endaligned
\end{eqnarray}
this leads naturally to a quantum channel Fisher information $J_{QC}(K_x)=\max_{\rho_{SA}}J_S(\rho_x)$ which is similarly related to the distance on quantum channels as
\begin{eqnarray}
\aligned
\label{eq:qcf}
J_{QC}(K_x)=&\lim_{dx\rightarrow 0}\frac{4B_{QC}^2(K_x,K_{x+dx})}{dx^2}\\
=&\lim_{dx\rightarrow 0}\frac{8[1-\cos \Theta_{QC}(K_x,K_{x+dx})]}{dx^2}\\
=&\lim_{dx\rightarrow 0}\frac{4\Theta_{QC}^2(K_x,K_{x+dx})}{dx^2}.
\endaligned
\end{eqnarray}
The quantum channel Fisher information quantifies the ultimate precision limit upon the optimization over the measurements and probe states
\begin{equation}
\label{eq:channelpre}
\delta \hat{x} \geq \frac{1}{\sqrt{nJ_{QC}(K_x)}}=\frac{1}{\sqrt{n}\lim_{dx\rightarrow 0}\frac{2\Theta_{QC}(K_x,K_{x+dx})}{|dx|}}.
\end{equation}
This connects the precision limit directly to the distance between quantum channels which provides a clear physical picture for the ultimate precision limit. The scaling of the ultimate precision limit can now be seen as a manifestation of the scaling of the distances between quantum channels as we now show. 
\begin{figure}[t]
\centering
\includegraphics[width=\linewidth]{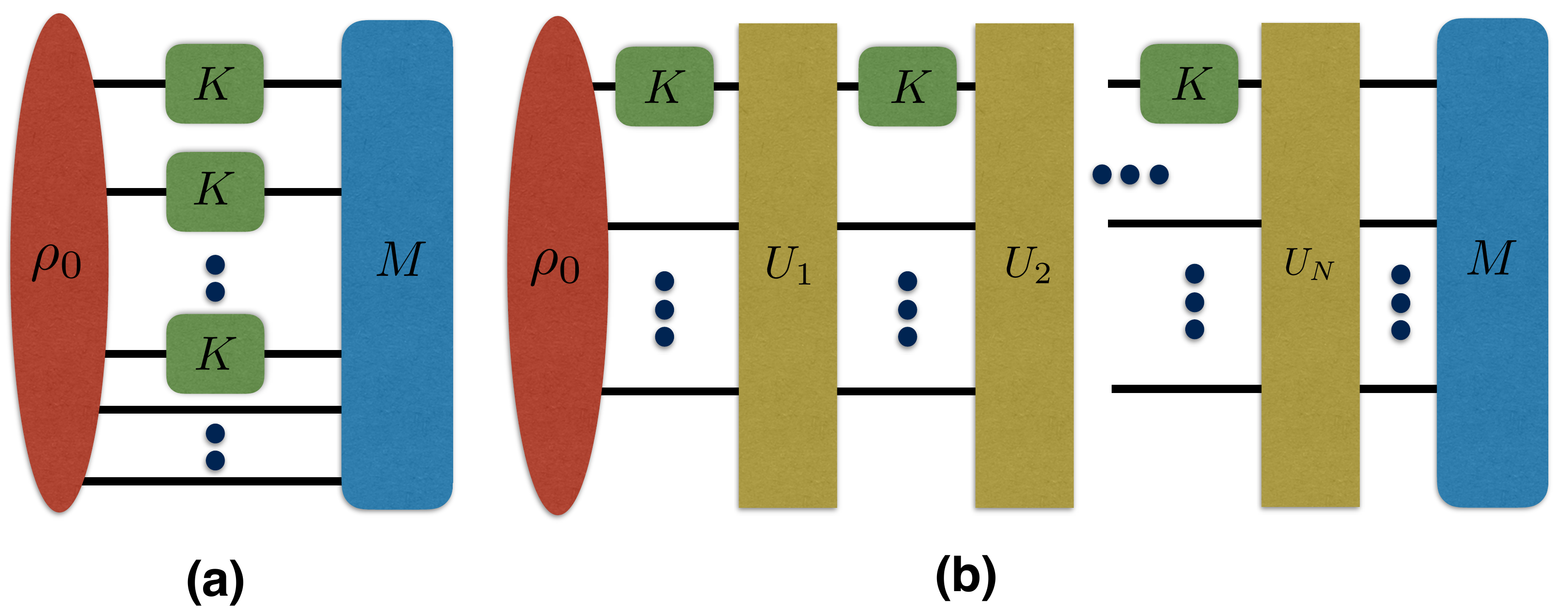}
\caption{ (a) Parallel Scheme. (b)Sequential Scheme.}
\label{fig:scheme}
\end{figure}

Two schemes on multiple uses of quantum channels are usually considered in quantum parameter estimation, the parallel scheme and the sequential scheme as shown in Fig.\ref{fig:scheme}. We will show that for both schemes, the scaling of the distances between two quantum channels are at most linear, which underlies the scaling for the Heisenberg limit.

\begin{figure}
\centering
\begin{minipage}{.35\textwidth}
  \centering
  \includegraphics[width=.9\linewidth]{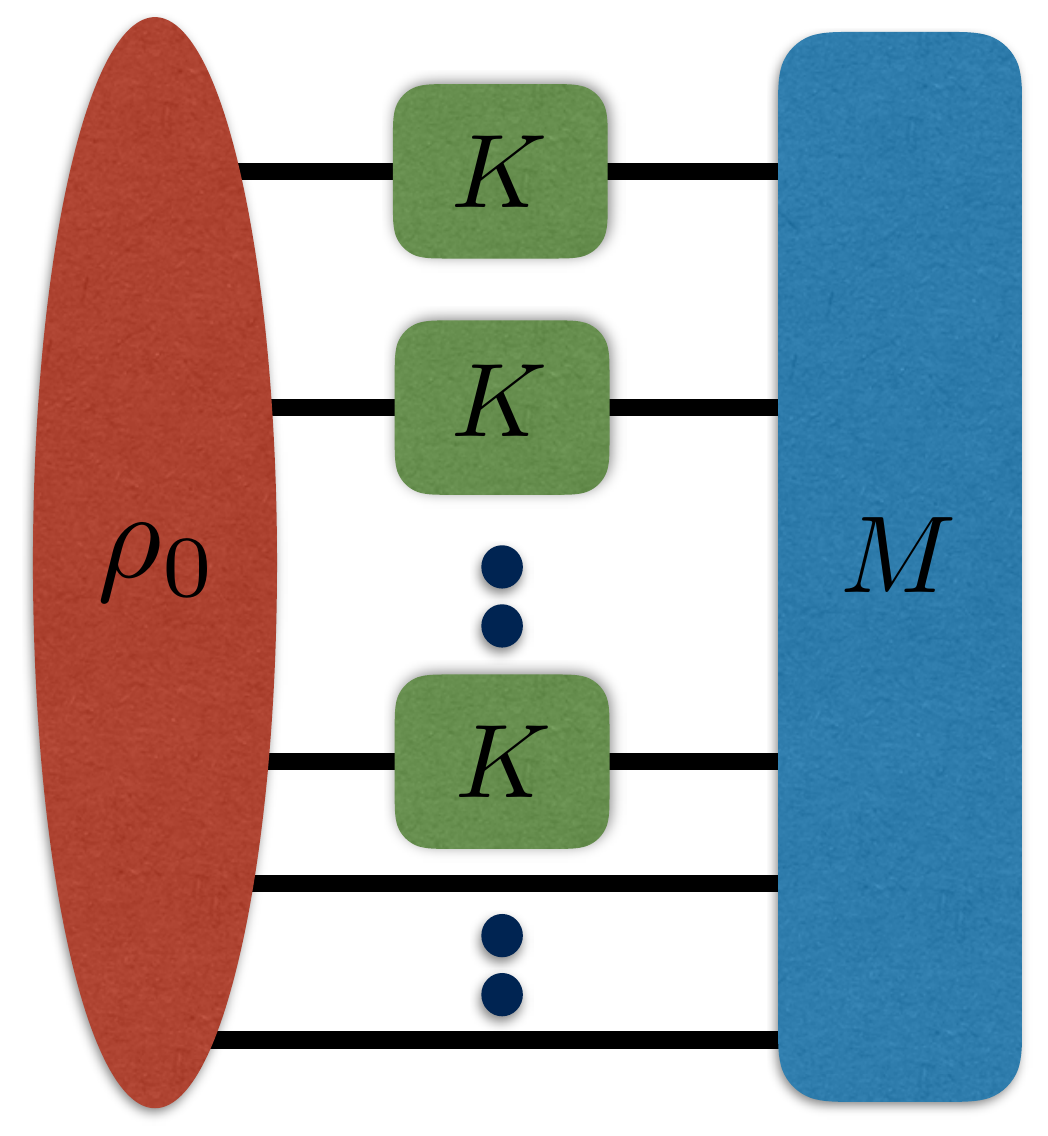}
  \caption{Parallel scheme with multiple uses of the channel.}
  \label{fig:parallel}
\end{minipage}%
\qquad
\begin{minipage}{.34\textwidth}
  \centering
  \includegraphics[width=.9\linewidth]{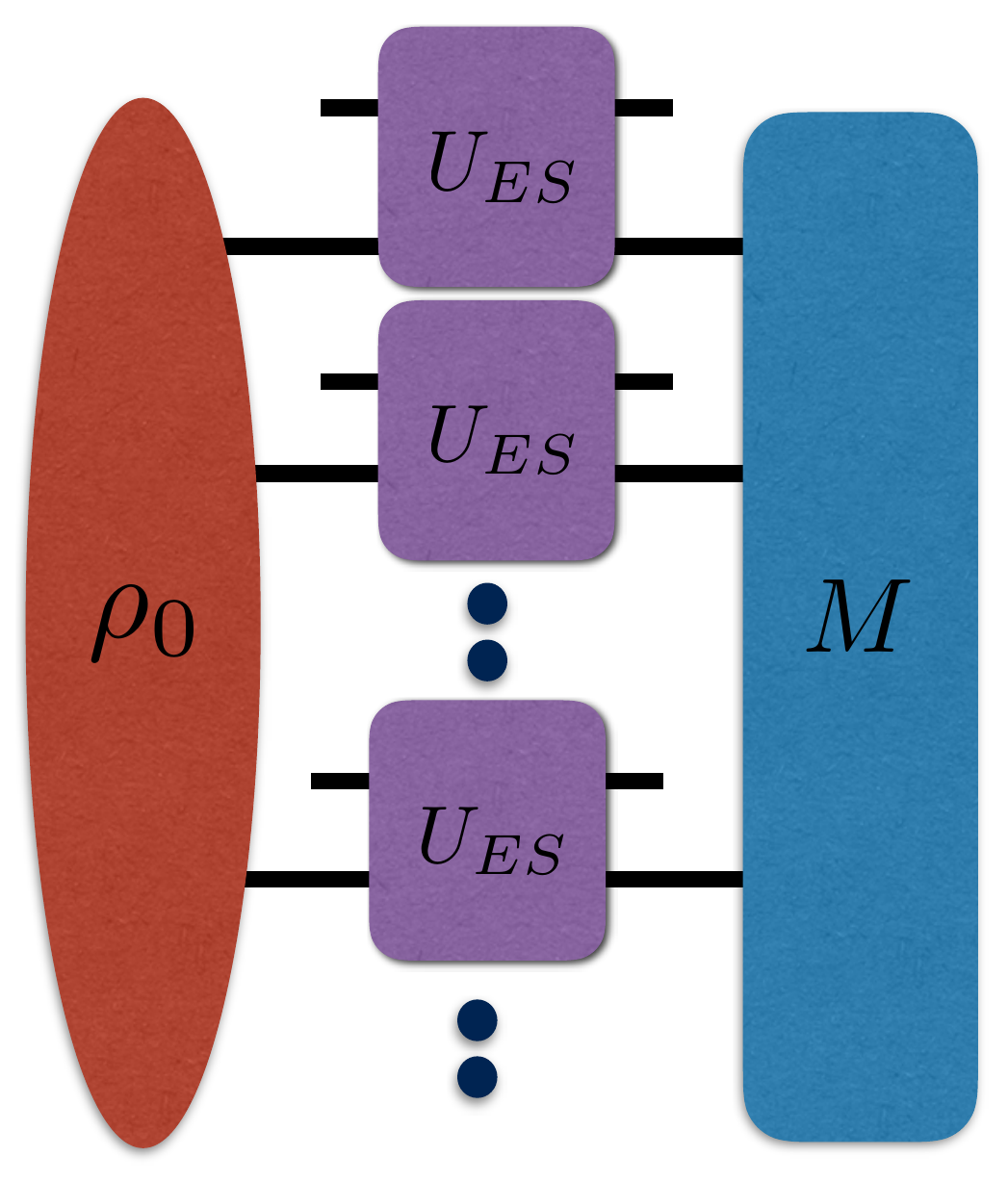}
  \caption{A unitary extension of the parallel scheme.}
  \label{fig:parallelu}
\end{minipage}
\end{figure}

For parallel scheme with $N$ uses of a  channel $K$ as shown in Fig.\ref{fig:parallel}, the total dynamics can be described by $K^{\otimes N}\otimes I_A$. If we denote $U_{ES}$ as one unitary extension of $K$, then $U_{ES}^{\otimes N}$ is a unitary extension of $K^{\otimes N}$ as shown in Fig.\ref{fig:parallelu}. Given two channels $K_1$ and $K_2$, we choose $U_{ES1}$ and $U_{ES2}$ as the unitary extension for $K_1$ and $K_2$ respectively which satisfies $\Theta_{QC}(K_1,K_2)=\Theta_{QC}(U_{ES1},U_{ES2})$. Now as $U_{ES1}^{\otimes N}$ and $U_{ES2}^{\otimes N}$ are unitary extensions of $K_1^{\otimes N}$ and $K_2^{\otimes N}$ respectively, we then have
\begin{eqnarray}
\label{eq:bound}
\aligned
\Theta_{QC}(K_1^{\otimes N},K_2^{\otimes N})&\leq \Theta_{QC}(U_{ES1}^{\otimes N}, U_{ES2}^{\otimes N}) \\
&=C[(U_{ES1}^\dagger U_{ES2})^{\otimes N}]\\
&\leq NC(U_{ES1}^\dagger U_{ES2})\\
&=N\Theta_{QC}(K_1,K_2).
\endaligned
\end{eqnarray}

\begin{figure}
\centering
\begin{minipage}{.7\textwidth}
  \centering
  \includegraphics[width=.9\linewidth]{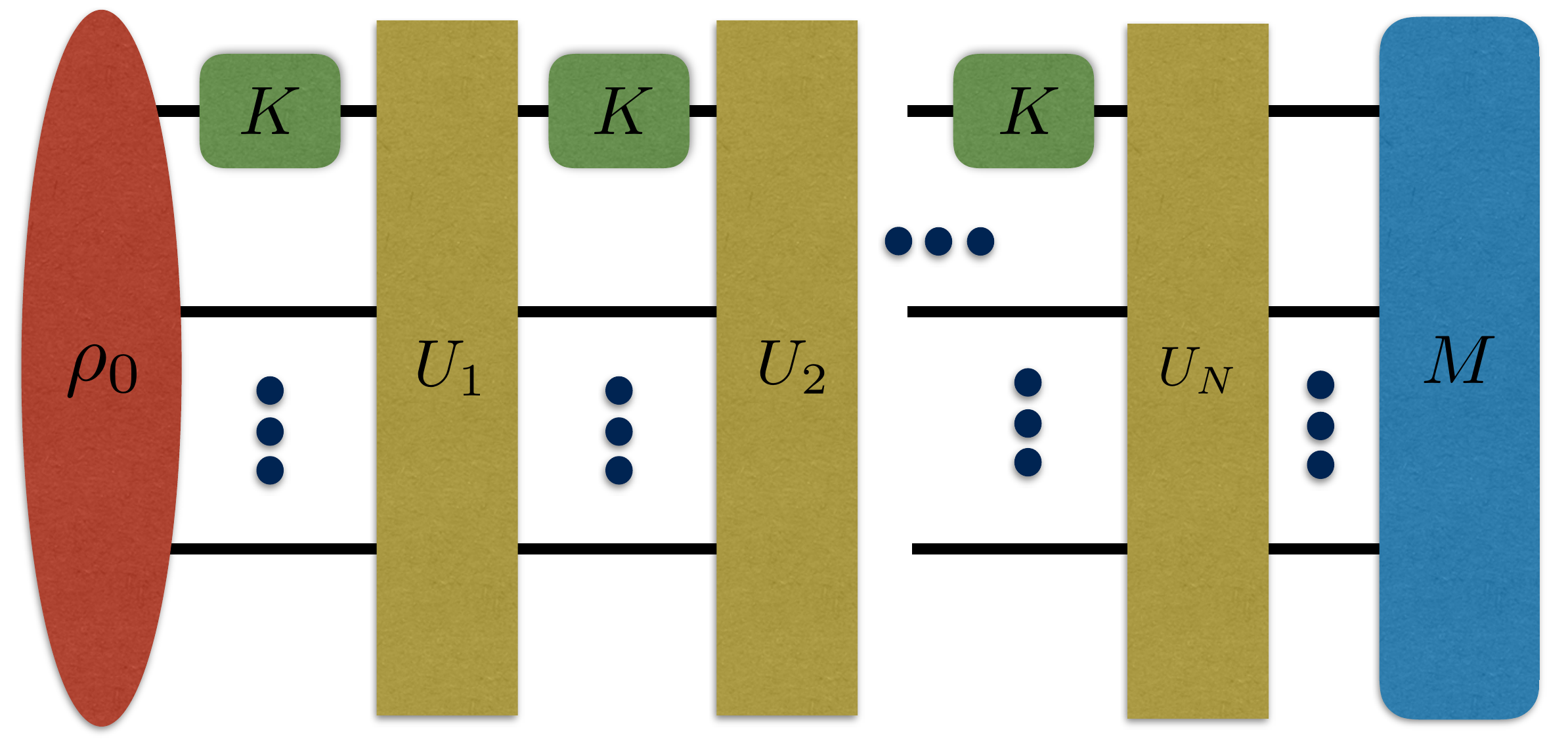}
  \caption{Sequential scheme with multiple uses of the channel.}
  \label{fig:seqsupp}
\end{minipage}%
\qquad
\begin{minipage}{.92\textwidth}
  \centering
  \includegraphics[width=.9\linewidth]{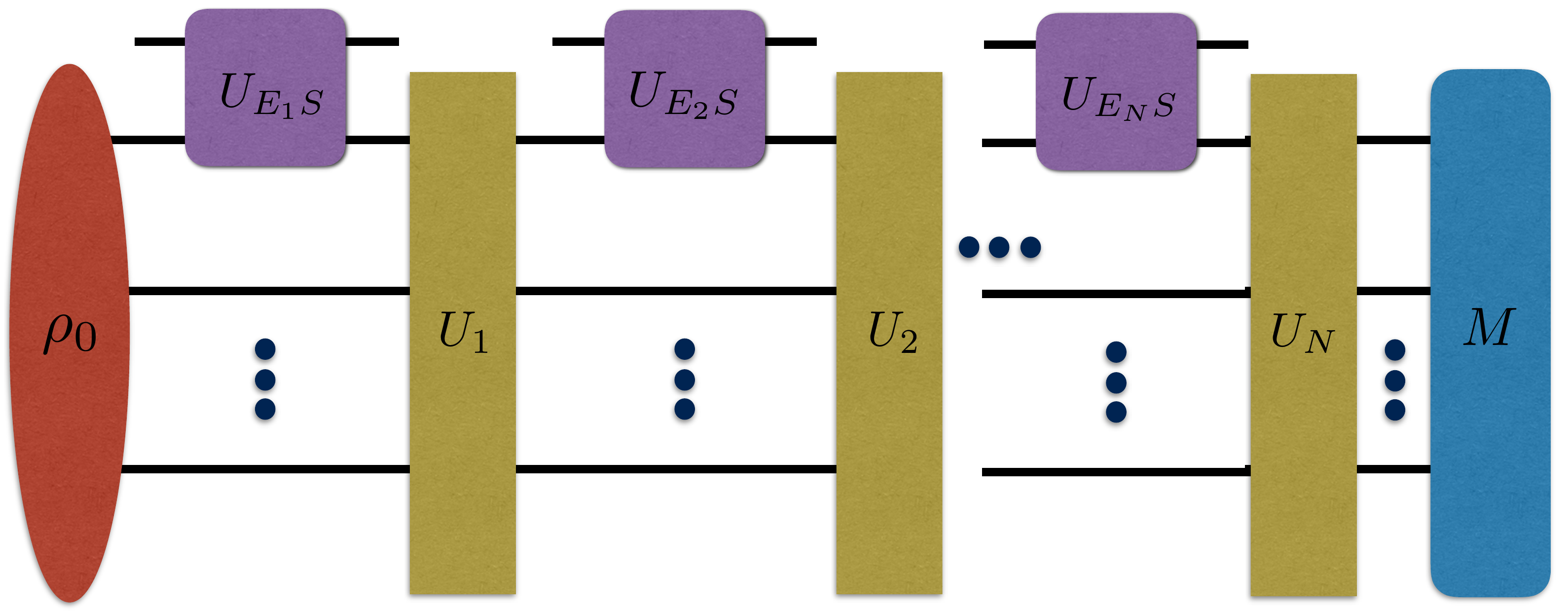}
  \caption{A unitary extension of the sequential scheme.}
  \label{fig:sequsupp}
\end{minipage}
\end{figure}

For the sequential scheme, we consider the general case that controls can be inserted between sequential uses of the channels. Any measurements that are used in the control can be substituted by controlled unitaries with ancillary systems, the controls interspersed between the channels can thus be taken as unitaries, which is shown in Fig.\ref{fig:seqsupp}. Parallel scheme can be seen as a special case of the sequential scheme by choosing the controls as SWAP gates on the system and different ancillary systems\cite{Rafal2014}. We show that with $N$ uses of the channel, the distance is still bounded above by $N\Theta_{QC}(K_1,K_2)$.

 We present the proof for the case of $N=2$, same line of argument works for general $N$.
For $N=2$, one unitary extension of $U_2K_1U_1K_1$ is $U_2U_{E_2S1}U_1U_{E_1S1}$, similarly $U_2U_{E_2S2}U_1U_{E_1S2}$ is a unitary extension of $U_2K_2U_1K_2$, here $U_{E_jSi}$ denote a unitary extension of $K_i$, $i=1,2$, with $E_j$ as the environment. We can choose $U_{E_jSi}$ such that
 $\Theta_{QC}(K_1,K_2)=\Theta_{QC}(U_{E_jS1},U_{E_jS2})$,
here all operators are understood as defined on the whole space so the multiplication makes sense, for example the control $U_1$, which only acts on the system and ancillaries, is understood as $U_1\otimes I_E$, an operator on the whole space including the environment. We then have
\begin{eqnarray}
\aligned
&\Theta_{QC}(U_2K_1U_1K_1,U_2K_2U_1K_2)\\
\leq& \Theta_{QC}(U_2U_{E_2S1}U_1U_{E_1S1}, U_2U_{E_2S2}U_1U_{E_1S2}) \\
=&C[   U_{E_1S1}^\dagger U_1^\dagger U_{E_2S1}^\dagger U_2^\dagger U_2U_{E_2S2}U_1U_{E_1S2}]\\
=&C[U_{E_1S1}^\dagger U_1^\dagger U_{E_2S1}^\dagger U_{E_2S2}U_1U_{E_1S2}]\\
=&C[(U_{E_1S1}^\dagger U_1^\dagger) (U_{E_2S1}^\dagger U_{E_2S2}) (U_1U_{E_1S1}) (U_{E_1S1}^\dagger U_{E_1S2})]\\
\leq&C[U_{E_2S1}^\dagger U_{E_2S2}]+C[U_{E_1S1}^\dagger U_{E_1S2}]\\
=& 2\Theta_{QC}(K_1,K_2),\\
\endaligned
\end{eqnarray}
i.e., with two uses of the channel, the distance is bounded above by $2\Theta_{QC}(K_1,K_2)$. With the same line of argument it is easy to show that with $N$ uses of the channel the distance is bounded above by $N\Theta_{QC}(K_1,K_2)$.

Substitute $K_1$ with $K_x$ and $K_2$ with $K_{x+dx}$, we have $\Theta_{QC}(NK_x,NK_{x+dx})\leq N\Theta_{QC}(K_x,K_{x+dx})$ for both schemes. From Eq.(\ref{eq:channelpre}) the ultimate precision limit is then bounded by
\begin{eqnarray}
\aligned
\label{eq:Heisenberg}
\delta \hat{x} 
&\geq \frac{1}{\lim_{dx\rightarrow 0}\frac{2\Theta_{QC}(K_x,K_{x+dx})}{\mid dx\mid}N\sqrt{n}} ,
\endaligned
\end{eqnarray}
 the scaling $1/N$ is called the Heisenberg scaling, which, as we showed, is just a manifestation of the fact that the distance between quantum channels can grow at most linearly with the number of channels. 

For $N$ uses of the channels under the parallel scheme we can also obtain a tighter bound as
\begin{eqnarray}
\label{eq:Nparallel}
\aligned
&2-2\cos \Theta_{QC}(K_1^{\otimes N},K_2^{\otimes N})\\
&\leq N\|2I-K_W-K_W^\dagger\|+N(N-1)\|I-K_W\|^2,
\endaligned
\end{eqnarray}
here $K_{W}=\sum_{i=1}^q\sum_{j=1}^qw_{ij}F_{1i}^\dagger F_{2j}$ as previously defined, and the inequality holds for any $W$ with $\|W\|\leq 1$
(see Appendix~\ref{app-upper-bound-parallel}).
In the asymptotical limit, $N(N-1)\|I-K_W\|^2$ is the dominating term, in that case we would like to choose a $W$ minimizing $\|I-K_W\|$ to get a tighter bound. This can be formulated as semi-definite programming with $\min_{\|W\|\leq 1} \|I-K_W\|=$
\begin{eqnarray}
\label{eq:sdpN}
\aligned
&\min \qquad t \\
s.t.\qquad &\left(\begin{array}{cc}
      I & W^\dagger  \\
      W & I \\
          \end{array}\right)\succeq 0,\\
      &\left(\begin{array}{cc}
      tI & (I-K_W)^\dagger  \\
      I-K_W & tI \\
          \end{array}\right)\succeq 0.
          \endaligned
          \end{eqnarray}

If we let $K_1=K_x$ and $K_2=K_{x+dx}$, then Eq.(\ref{eq:Nparallel}) provides bounds on the scalings in quantum parameter estimation, which is consistent with the studies in quantum metrology\cite{Fujiwara2008,Escher2011,Rafal2012,Jan2013,Rafal2014} but here with a more general context
(see also Ref.~\cite{Yuan2017npj}).

Next we show how the tools unify quantum parameter estimation and the perfect quantum channel discrimination\cite{Acin01,Duan2007,Duan2008,Cheng2012, ChiribellaDP08,DuanFY09,Harrow2010}.

Given two quantum channels $K_1$ and $K_2$, they can be perfectly discriminated with one use of the channels if and only if there exists a $\rho_{SA}$ such that $K_1\otimes I_A(\rho_{SA})$ and $K_2\otimes I_A(\rho_{SA})$ are orthogonal, i.e., $\min_{\rho_{SA}}F_S[K_1\otimes I_A(\rho_{SA}),K_2\otimes I_A(\rho_{SA})]=0$, which is the same as $\Theta_{QC}(K_1,K_2)=\frac{\pi}{2}$. When $K_1$ and $K_2$ can not be perfectly discriminated with one use of the channel, finite number of uses may able to achieve the task\cite{DuanFY09}. This is in contrast to the perfect discrimination of non-orthogonal states which always requires infinite number of copies. The minimum number of uses needed for perfect channel discrimination should satisfy $\Theta_{QC}(NK_1,NK_2)=\frac{\pi}{2}$. The perfect channel discrimination is thus determined by the distances between quantum channels, and the scalings of $\Theta_{QC}(NK_1,NK_2)$ obtained before can be used to determine the minimum $N$. For example, from $\Theta_{QC}(NK_1,NK_2)\leq N\Theta_{QC}(K_1,K_2)$ we can obtain a lower bound on $N$ as
\begin{equation}
\label{eq:lowerN}
N\geq \lceil\frac{\pi}{2\Theta_{QC}(K_1,K_2)}\rceil,
\end{equation}
where $\lceil x\rceil$ is the smallest integer not less than $x$. This bound is tighter than existing bounds for noisy channels\cite{Cheng2012} and for unitary channels it reduces to the formula which is known to be tight\cite{Acin01}. For noisy channels under the parallel scheme we can also substitute $\Theta_{QC}(K_1^{\otimes N},K_2^{\otimes N})=\frac{\pi}{2}$ into the inequality (\ref{eq:Nparallel}) to get a tighter bound.

The lower bound on minimum $N$ can also be obtained via a connection to quantum metrology. Given two channels $K_1$ and $K_2$, let $K_x,  x\in [a,b]$ as a path connecting $K_1$ and $K_2$. With $N$ uses of the channel under the parallel strategy we have $\sqrt{J_{QC}(K_x^{\otimes N})}=\lim_{dx\rightarrow 0}\frac{2\Theta_{QC}(K_x^{\otimes N},K_{x+dx}^{\otimes N})}{\mid dx\mid}$. From the triangular inequality
\begin{eqnarray}
\label{eq:connect}
\aligned
\Theta_{QC}(K_1^{\otimes N},K_2^{\otimes N})&\leq \int_a^b \lim_{dx\rightarrow 0}\frac{\Theta_{QC}(K_x^{\otimes N},K_{x+dx}^{\otimes N})}{dx} dx\\
&=\frac{1}{2}\int_a^b \sqrt{J_{QC}(K_x^{\otimes N})} dx.
\endaligned
\end{eqnarray}
This connects the prefect channel discrimination to the ultimate precision limit. By choosing different paths various useful lower bounds on the minimum number of uses for perfect channel discrimination can be obtained.

For example, given $K_0(\rho)=e^{i\theta\sigma_1}\rho e^{-i\theta\sigma_1}$ and $K_1=\frac{1+\eta}{2}\rho+ \frac{1-\eta}{2}\sigma_3 \rho\sigma_3$, where $\sigma_1, \sigma_2$ and $\sigma_3$
          are Pauli matrices and assume $\theta=0.3, \eta=0.5$. For the parallel strategy the lower bound given by Eq.(\ref{eq:lowerN}) is $N\geq \lceil\frac{\pi}{2\Theta_{QC}(K_0,K_1)}\rceil=3$.
          If we choose a simple path $K_x=(1-x)K_0+xK_1$, $x\in[0,1]$, which is a line segment connecting $K_0$ to $K_1$, then with the connection provided by Eq.(\ref{eq:connect}) we obtain $N\geq 4$. Other paths may be explored to further improve the bound.
          By using the inequality (\ref{eq:Nparallel}) with the $W$ obtained from the semi-definite programming that minimizes $\|I-K_W\|$, we get $N\geq 5$. For any $N$ we can also choose the $W$ to minimize $N\|2I-K_W-K_W^\dagger\|+N(N-1)\|I-K_W\|^2$, it turns out the minimum $N$ such that $\min_{\|W\|\leq 1} N\|2I-K_W-K_W^\dagger\|+N(N-1)\|I-K_W\|^2\geq 2$ is $6$, thus $N\geq 6$. For comparison we also explicitly computed the actual distance $\Theta_{QC}(K_0^{\otimes N},K_1^{\otimes N})$ with the increasing of $N$, it turns out that the minimum $N$ such that $\Theta_{QC}(K_0^{\otimes N},K_1^{\otimes N})=\frac{\pi}{2}$ is actually $6$. All computations here are done with the CVX package in Matlab\cite{CVX}.

\section{Summary}
A fidelity function defined directly on quantum channels is provided, which leads to various distance measures defined directly on quantum channels, as well as a new Fisher information on quantum channel. This forms another hierarchy for fidelity functions and Fisher information as shown in the table:
\begin{displaymath}
\xymatrix{
  F_C(p_1,p_2) \ar[d] \ar[r] & F_S(\rho_1,\rho_2) \ar[d] \ar[r] & F_{QC}(K_1,K_2) \ar[d] \\
  \Theta_C(p_1,p_2) \ar[d] \ar[r] & \Theta_S(\rho_1,\rho_2) \ar[d] \ar[r] & \Theta_{QC}(K_1,K_2) \ar[d] \\
  J_C[p(y|x)] \ar[r] & J_S(\rho_x) \ar[r] & J_{QC}(K_x)   }
\end{displaymath}
where $\cos \Theta_i=F_i$ and $J_i=\lim_{dx\rightarrow 0}\frac{4\Theta_i^2}{dx^2}$, $i\in \{C,S,QC\}$. In this table the functions on quantum states equal to the optimized value over all measurements of the corresponding functions on probability distribution, and the functions on quantum channels equal to the optimized value over all probe states of the corresponding functions on quantum states. This framework connects quantitatively the ultimate precision limit and the distance between quantum channels, which provided a clear physical picture for the ultimate precision limit in quantum metrology. It also provide a unified framework for the continuous case in quantum parameter estimation and the discrete case in perfect quantum channel discrimination, with this framework the progress in one field can then be readily used to stimulate the progress of the other field. We expect these tools will find wide applications in many other fields of quantum information science.

\appendix

\section{Formula to compute $\Theta_{QC}(K_1,K_2)$ }
\label{app-compute}
We show that the distance between two quantum channels $\Theta_{QC}(K_1, K_2)=\min_{\{U_{ES1},U_{ES2}\}}\Theta_{QC}(U_{ES1}, U_{ES2})$ can be computed from the Kraus operators of $K_1$ and $K_2$ as $\Theta_{QC}(K_1, K_2)=\min_{\{U_{ES1},U_{ES2}\}}\Theta_{QC}(U_{ES1}, U_{ES2})=\arccos \max_{|W|\leq 1} \frac{1}{2}\lambda_{\min}[K_W+K_W^\dagger],$ here $U_{ESi}$ are unitary extensions of $K_i$, $i\in\{1,2\}$ and $\lambda_{\min}(K_W+K^\dagger_W)$ denotes the minimum eigenvalue of $K_W+K^\dagger_W$ with $K_W=\sum_{ij}w_{ij}F_{1i}^\dagger F_{2j}$, $F_{1i}$, $F_{2j}$ denotes the Kraus operators of $K_1$ and $K_2$, $w_{ij}$ denotes the $ij$-th entry of a $q\times q$ matrix $W$ with $\|W\|\leq 1$($\|\cdot\|$ is the operator norm which equals to the maximum singular value), $q$ is the number of the Kraus operators. Furthermore the minimization on both $U_{ES1}$ and $U_{ES2}$ can be reduced to the minimization of just one
\begin{eqnarray}
\aligned
\Theta_{QC}(K_1, K_2)&=\min_{\{U_{ES1},U_{ES2}\}}\Theta_{QC}(U_{ES1}, U_{ES2})\\
&=\min_{U_{ES1}}\Theta_{QC}(U_{ES1}, U_{ES2})\\
&=\min_{U_{ES2}}\Theta_{QC}(U_{ES1}, U_{ES2}).\\
\endaligned
\end{eqnarray}

We start by a general unitary extension for any given channel $K(\rho)=\sum_{j=1}^q F_j\rho F^\dagger_j$ with
$\sum_{j=1}^q F^\dagger_jF_j=I$, which maps from a $m_1$- to $m_2$- dimensional Hilbert space,
\begin{align}
\label{eqn-U-general-form0}
U_{ES}=(W_E \otimes I_{m_2})
\underbrace{
\begin{bmatrix}
F_1 & * & * & \cdots & *
\\
F_2 & * & * & \cdots & *
\\
\vdots & &\vdots & & \vdots
\\
F_{q} & * & * & \cdots & *\\
0 & * & * & \cdots & *\\
\vdots & &\vdots & & \vdots\\
0 & * & * & \cdots & *\\
\end{bmatrix}
}_{\displaystyle U},
\end{align}
where $W_E \in U(p)$ only acts on the environment and can be chosen arbitrarily, here $U(p)$ denotes the set of $p\times p$ unitary operators with $p\geq q$ as $p-q$ zero Kraus operators can be added. Here only the first $m_1$ columns of $U$ are fixed, the freedom of other columns can be represented as
\begin{align}
U_{ES}=(W_E \otimes I_{m_2})
\begin{bmatrix}
F_1 & * & * & \cdots & *
\\
F_2 & * & * & \cdots & *
\\
\vdots & &\vdots & & \vdots
\\
F_{q} & * & * & \cdots & *\\
0 & * & * & \cdots & *\\
\vdots & &\vdots & & \vdots\\
0 & * & * & \cdots & *\\
\end{bmatrix}
\begin{bmatrix}
I_{m_1} & 0\\
0 & V
\end{bmatrix}
\end{align}
where $V$ can be any unitary.

For two channels $K_1$ and $K_2$, with $K_1(\rho)=\sum_{j=1}^q F_{1j}\rho F^\dagger_{1j}$ and $K_2(\rho)=\sum_{j=1}^q F_{2j}\rho F^\dagger_{2j}$, the unitary extensions can be written as
\begin{align}
\label{eqn-U-general-form1}
U_{ES1}=(W_{E1} \otimes I_{m_2})
\begin{bmatrix}
F_{11} & * & * & \cdots & *
\\
F_{12} & * & * & \cdots & *
\\
\vdots & &\vdots & & \vdots
\\
F_{1q} & * & * & \cdots & *\\
0 & * & * & \cdots & *\\
\vdots & &\vdots & & \vdots\\
0 & * & * & \cdots & *\\
\end{bmatrix}\begin{bmatrix}
I_{m_1} & 0\\
0 & V_1
\end{bmatrix},
\end{align}

\begin{align}
U_{ES2}=(W_{E2} \otimes I_{m_2})
\begin{bmatrix}
F_{21} & * & * & \cdots & *
\\
F_{22} & * & * & \cdots & *
\\
\vdots & &\vdots & & \vdots
\\
F_{2q} & * & * & \cdots & *\\
0 & * & * & \cdots & *\\
\vdots & &\vdots & & \vdots\\
0 & * & * & \cdots & *\\
\end{bmatrix}
\begin{bmatrix}
I_{m_1} & 0\\
0 & V_2
\end{bmatrix},
\end{align}
then $$U_{ES1}^\dagger U_{ES2}=\begin{bmatrix}
I_{m_1} & 0\\
0 & V_1^\dagger
\end{bmatrix}
\begin{bmatrix}
K_W & * & * & \cdots & *
\\
* & * & * & \cdots & *
\\
\vdots & &\vdots & & \vdots
\\
* & * & * & \cdots & *\\
* & * & * & \cdots & *\\
\vdots & &\vdots & & \vdots\\
* & * & * & \cdots & *\\
\end{bmatrix}\begin{bmatrix}
I_{m_1} & 0\\
0 & V_2
\end{bmatrix},$$ here $K_W=\sum_{ij}w_{ij}F_{1i}^\dagger F_{2j}$, where $w_{ij}$ is the $ij$-th entry of $W$, here $W$ is the first $q\times q$ block of $W_{E1}^\dagger W_{E2}$, i.e.,
$W_{E1}^\dagger W_{E2}
=\begin{bmatrix}
W & *\\
* & *
\end{bmatrix}$. It is easy to see that $\|W\|\leq 1$, conversely for any $W$ with $\|W\|\leq 1$ it can be imbedded as the first $q\times q$ block of a unitary matrix\cite{Choi}. Thus by varying $W_{E1}$ and $W_{E2}$ we can take $W$ to be any $q\times q$ matrix with $\|W\|\leq 1$. $\min_{U_{ES1},U_{ES2}}\Theta_{QC}(U_{ES1},U_{ES2})=\min_{U_{ES1},U_{ES2}}C(U_{ES1}^\dagger U_{ES2})$ is now reduced to the optimizing over $V_1$, $V_2$ and $W$.

First note that for a fixed $W$, the first block of $U_{ES1}^\dagger U_{ES2}$ is always $K_W$, as $\begin{bmatrix}
I_{m_1} & 0\\
0 & V_1^\dagger
\end{bmatrix}$ and $\begin{bmatrix}
I_{m_1} & 0\\
0 & V_2
\end{bmatrix}$ do not change the first block. It has been shown in \cite{Fung3} that for any unitary that has $K_W$ as the first block
$$U=\begin{bmatrix}
K_W & * & * & \cdots & *
\\
* & * & * & \cdots & *
\\
\vdots & &\vdots & & \vdots
\\
* & * & * & \cdots & *\\
* & * & * & \cdots & *\\
\vdots & &\vdots & & \vdots\\
* & * & * & \cdots & *\\
\end{bmatrix},$$  $\|U\|_{\max} \geq \arccos [\frac{1}{2}\lambda_{\min}(K_W+K_W^\dagger)]$, where $\|U\|_{\max}$ is defined in Eq.(1) of the main text. Thus we have $\| U_{ES1}^\dagger U_{ES2}\|_{\max} \geq \arccos [\frac{1}{2}\lambda_{\min}(K_W+K_W^\dagger)]$. What's more it was also shown that there exists a unitary $V_2$ with $$U_{V_2}=\begin{bmatrix}
K_W & * & * & \cdots & *
\\
* & * & * & \cdots & *
\\
\vdots & &\vdots & & \vdots
\\
* & * & * & \cdots & *\\
* & * & * & \cdots & *\\
\vdots & & \vdots& & \vdots\\
* & * & * & \cdots & *\\
\end{bmatrix}\begin{bmatrix}
I_{m_1} & 0\\
0 & V_2
\end{bmatrix}$$ such that $\|U_{V_2}\|_{\max}=\arccos [\frac{1}{2}\lambda_{\min}(K_W+K_W^\dagger)]$ achieves the bound\cite{Fung3}.
Similarly the bound can also be achieved by exploring the freedom in rows, i.e., there exists a unitary $V_1$, $$U_{V_1}=\begin{bmatrix}
I_{m_1} & 0\\
0 & V_1^\dagger
\end{bmatrix}\begin{bmatrix}
K_W & * & * & \cdots & *
\\
* & * & * & \cdots & *
\\
\vdots & &\vdots & & \vdots
\\
* & * & * & \cdots & *\\
* & * & * & \cdots & *\\
\vdots & & \vdots& & \vdots\\
* & * & * & \cdots & *\\
\end{bmatrix},$$ such that $\|U_{V_1}\|_{\max}=\arccos [\frac{1}{2}\lambda_{\min}(K_W+K_W^\dagger)].$ Thus for a fixed $W$, $\min_{\{V_1,V_2\}} \|U^\dagger_{ES1}U_{ES2}\|_{\max}=\arccos [\frac{1}{2}\lambda_{\min}(K_W+K_W^\dagger)]$.

Next we optimize over $W$. Basically we need to find $W$ such that $\arccos \frac{1}{2}\lambda_{\min}[K_W+K_W^\dagger]$ is minimized, which is equivalent to find $\max_{|W|\leq 1}\frac{1}{2}\lambda_{\min}[K_W+K_W^\dagger]$. Note that the freedom of global phase from $\|\cdot\|_{\max}$ to $\|\cdot\|_g$(see main text for definitions) has been included in the freedom of $W$ and since $\max_{|W|\leq 1}\frac{1}{2}\lambda_{\min}[K_W+K_W^\dagger]\geq \frac{1}{2}\lambda_{\min}[K_{\mathbf{0}}+K_{\mathbf{0}}^\dagger]=0$, we have $\arccos \max_{|W|\leq 1}\frac{1}{2}\lambda_{\min}[K_W+K_W^\dagger]\leq \frac{\pi}{2}$. Thus $\min_{U_{ES1},U_{ES2}}C(U_{ES1}^\dagger U_{ES2})=\arccos \max_{|W|\leq 1}\frac{1}{2}\lambda_{\min}[K_W+K_W^\dagger]$, i.e.
\begin{eqnarray}
\label{eq:thetaQC}
\aligned
\Theta_{QC}(K_1,K_2)&=\min_{U_{ES1},U_{ES2}}\Theta_{QC}(U_{ES1},U_{ES2})\\
&= \arccos \max_{|W|\leq 1}\frac{1}{2}\lambda_{\min}[K_W+K_W^\dagger].
\endaligned
\end{eqnarray}
It is obvious that the freedom of $W$ can be achieved by only varying $W_1$ or $W_2$, thus the equality can be attained by just exploring the freedom of $V_1$ and $W_1$, or $V_2$ and $W_2$. We then have
\begin{eqnarray}
\aligned
&\min_{U_{ES1},U_{ES2}}\Theta_{QC}(U_{ES1},U_{ES2})\\
&=\min_{U_{ES1}}\Theta_{QC}(U_{ES1},U_{ES2})\\
&=\min_{U_{ES2}}\Theta_{QC}(U_{ES1},U_{ES2})\\
&=\arccos \max_{|W|\leq 1}\frac{1}{2}\lambda_{\min}[K_W+K_W^\dagger].
\endaligned
\end{eqnarray}

Next we show that this distance measure has a connection to the minimum distance between equivalent Kraus operators.  Given two quantum channels, $K_1(\rho_S)=\sum_{i=1}^q F_{1i}\rho_S F^\dagger_{1i}$ and $K_2(\rho_S)=\sum_{i=1}^q F_{2i}\rho_S F^\dagger_{2i}$(zero Kraus operators can be appended if the number of the Kraus operators are not the same), 
by appending additional $p-q$ zero Kraus operators, we have the Kraus operators for $K_1$ and $K_2$ as $\{F_{11},F_{12},\cdots, F_{1q},0,\cdots,0\}$ and $\{F_{21},F_{22},\cdots, F_{2q},0,\cdots,0\}$ respectively. Equivalent Kraus operators for $K_1$ and $K_2$ can be represented as $\tilde{F}_{1i}=\sum_k u_{ik}F_{1k}$ and $\tilde{F}_{2i}=\sum_k v_{ik}F_{2k}$ where $u_{ik}$ and $v_{ik}$ are entries of $U,V\in U(p)$ respectively, here $1\leq i\leq p$. Then
\begin{eqnarray}
\aligned
&\min_{\{\tilde{F}_{1i}\},\{\tilde{F}_{2i}\}} \| \sum_{i=1}^p (\tilde{F}_{1i}-\tilde{F}_{2i})^\dagger (\tilde{F}_{1i}-\tilde{F}_{2i})\| \\
=&\min_{\{\tilde{F}_{1i}\},\{\tilde{F}_{2i}\}}\| 2I-\sum_{i=1}^p (\tilde{F}_{1i}^\dagger \tilde{F}_{2i}+\tilde{F}_{2i}^\dagger \tilde{F}_{1i}\|\\
=&\min_{W} [2-\lambda_{\min}(K_W+K_W^\dagger)],
\endaligned
\end{eqnarray}
where $K_W=\sum_{ij}^q w_{ij} F_{1i}^\dagger F_{2j}$ and $w_{ij}$ is the $ij$-th entry of $W$, which is the first $q\times q$ block of $U^\dagger V$ and can be any $q\times q$ matrix with $\|W\|\leq 1$ by varying $U$ and $V$, i.e., by varying the equivalent representations of $K_1$ and $K_2$. Thus
\begin{eqnarray}
\aligned
&\min_{\{\tilde{F}_{1i}\},\{\tilde{F}_{2i}\}} \| \sum_{i=1}^p (\tilde{F}_{1i}-\tilde{F}_{2i})^\dagger (\tilde{F}_{1i}-\tilde{F}_{2i})\| \\
=&\min_{\|W\|\leq 1} [2-\lambda_{\min}(K_W+K_W^\dagger)]\\
=&2-\max_{\|W\|\leq 1}\lambda_{\min}(K_W+K_W^\dagger)\\
=&2-2\cos \Theta_{QC}(K_1,K_2),
\endaligned
\end{eqnarray}
we then have
\begin{eqnarray}
\aligned
B_{QC}^2(K_1,K_2)=2-2\cos \Theta_{QC}(K_1,K_2)=\min_{\{\tilde{F}_{1i}\},\{\tilde{F}_{2i}\}} \| \sum_{i=1}^p (\tilde{F}_{1i}-\tilde{F}_{2i})^\dagger (\tilde{F}_{1i}-\tilde{F}_{2i})\|.
\endaligned
\end{eqnarray}

\section{$\Theta_{QC}(K_1,K_2)$ defines a metric on quantum channels}
\label{app-metric}
We show that $\Theta_{QC}(K_1,K_2)$ defines a metric on quantum channels.

First we show that $\Theta_{QC}(U_1,U_2)=C(U_1^\dagger U_2)$, where $C$ is defined in the main text, is a metric on unitary channels.

We start by listing some useful properties of $C(U)$:
 \begin{eqnarray}
 \aligned
 C(V^\dagger U V)&=C(U);\\
 C(U_1\otimes U_2)&\leq C(U_1)+C(U_2);\\
 C(U_1U_2)&\leq C(U_1)+C(U_2);\\
 \endaligned
 \end{eqnarray}
where $V$ is any unitary operator. The first equality is obvious from the definition; the second inequality can be easily verified using the formula $C(U)=\frac{\theta_{\max}-\theta_{\min}}{2}$ when $\theta_{\max}-\theta_{\min}\leq \pi$, the equality is saturated when $C(U_1)+C(U_2)\leq \frac{\pi}{2}$; proof of the third inequality can be found in \cite{ChildsPR00,YuanUniversal}.

It is obvious that $\Theta_{QC}(U, U)=0$ and $\Theta_{QC}(U_1,U_2)=\Theta_{QC}(U_2,U_1)>0$ if $U_1\neq U_2$. 
And since
\begin{eqnarray}
\aligned
\Theta_{QC}(U_1,U_3)&=C(U_1^\dagger U_3)\\
&=C(U_1^\dagger U_2 U_2^\dagger U_3)\\
&\leq C(U_1^\dagger U_2)+C(U_2^\dagger U_3)\\
&=\Theta_{QC}(U_1,U_2)+\Theta_{QC}(U_2,U_3).
\endaligned
\end{eqnarray}
where for the inequality we have used the property that $C(U_1U_2)\leq C(U_1)+C(U_2)$. This shows that $\Theta_{QC}(U_1,U_2)$ is a metric on unitary operators.

For two general channels, $\Theta_{QC}(K_1, K_2)=\min_{U_{ES1}}\Theta_{QC}(U_{ES1}, U_{ES2})=\min_{U_{ES2}}\Theta_{QC}(U_{ES1}, U_{ES2})$ where $U_{ES1}$ and $U_{ES2}$ are unitary extensions for $K_1$ and $K_2$ respectively.
It is easy to see that $\Theta_{QC}(K_1,K_2)=\Theta_{QC}(K_2,K_1)\geq0$ and the equality is saturated only when $K_1=K_2$. We show that $\Theta_{QC}$ also satisfies the triangular inequality as
\begin{eqnarray}
\aligned
&\Theta_{QC}(K_1, K_3)\\
=&\min_{U_{ES1}}\Theta_{QC}(U_{ES1}, U_{ES3})\\
=&\min_{U_{ES1}}C(U_{ES1}^\dagger U_{ES3})\\
=&\min_{U_{ES1}}C(U_{ES1}^\dagger U_{ES2}U_{ES2}^\dagger U_{ES3})\\
\leq& \min_{U_{ES1}}[C(U_{ES1}^\dagger U_{ES2})+C(U_{ES2}^\dagger U_{ES3})]
\endaligned
\end{eqnarray}
the last inequality is valid for any $U_{ES2}$, specially we can choose the $U_{ES2}$ which minimizes $C(U_{ES2}^\dagger U_{ES3})$, thus
\begin{eqnarray}
\aligned
\Theta_{QC}(K_1, K_3)&\leq \min_{U_{ES1}}C(U_{ES1}^\dagger U_{ES2})+\Theta_{QC}(K_2,K_3)\\
&=\Theta_{QC}(K_1,K_2)+\Theta_{QC}(K_2,K_3).
\endaligned
\end{eqnarray}
$\Theta_{QC}(K_1,K_2)$ thus defines a metric on the space of quantum channels.

\section{Upper bound of the distance with $N$ parallel channels}
\label{app-upper-bound-parallel}

Given two quantum channels, $K_1(\rho_S)=\sum_{i=1}^q F_{1i}\rho_S F^\dagger_{1i}$ and $K_2(\rho_S)=\sum_{i=1}^q F_{2i}\rho_S F^\dagger_{2i}$, 
by appending $p-q$ zero Kraus operators, we have the Kraus operators for $K_1$ and $K_2$ as $\{F_{11},F_{12},\cdots, F_{1q},0,\cdots,0\}$ and $\{F_{21},F_{22},\cdots, F_{2q},0,\cdots,0\}$ respectively. All the equivalent Kraus operators for $K_1$ and $K_2$ can be represented as $\tilde{F}_{1i}=\sum_k u_{ik}F_{1k}$ and $\tilde{F}_{2i}=\sum_k v_{ik}F_{2k}$ where $u_{ik}$ and $v_{ik}$ are entries of $U,V\in U(p)$ respectively, here $1\leq i\leq p$.

With $N$ channels in parallel, one representation of the Kraus operators for $K_1^{\otimes N}$ can be written as $\tilde{F}_{1i_1,1i_2,\cdots, 1i_N}=\tilde{F}^{(1)}_{1i_1}\otimes \tilde{F}^{(2)}_{1i_2}\otimes\cdots \otimes \tilde{F}^{(N)}_{1i_N}$, similarly for $K_2^{\otimes N}$ we have $\tilde{F}_{2i_1,2i_2,\cdots, 2i_N}=\tilde{F}^{(1)}_{2i_1}\otimes \tilde{F}^{(2)}_{2i_2}\otimes\cdots \otimes \tilde{F}^{(N)}_{2i_N}$, where $\tilde{F}^{(l)}_{1i_l}=\sum_{k=1}^p u_{i_lk}F_{1k}$ are one Kraus operators of the $l$-th channel of $K_1^{\otimes N}$, similarly $\tilde{F}^{(l)}_{2i_l}=\sum_{k=1}^p v_{i_lk}F_{2k}$ are one Kraus operators of the $l$-th channel of $K_2^{\otimes N}$. As $\{\tilde{F}_{1i_1,1i_2,\cdots, 1i_N}\}$ and $\{\tilde{F}_{2i_1,2i_2,\cdots, 2i_N}\}$ are just one particular Kraus representation of $K_1^{\otimes N}$ and $K_2^{\otimes N}$ respectively, we then have
\begin{widetext}
\begin{eqnarray}
\nonumber
\aligned
2-2\cos \Theta_{QC}(K_1^{\otimes N},K_2^{\otimes N})\leq \|\sum_{i_1,i_2,\cdots, i_N}(\tilde{F}_{1i_1,1i_2,\cdots, 1i_N}-\tilde{F}_{2i_1,2i_2,\cdots, 2i_N})^\dagger(\tilde{F}_{1i_1,1i_2,\cdots, 1i_N}-\tilde{F}_{2i_1,2i_2,\cdots, 2i_N})\|,
\endaligned
\end{eqnarray}
since
\begin{eqnarray}
\aligned
&\tilde{F}_{1i_1,1i_2,\cdots, 1i_N}-\tilde{F}_{2i_1,2i_2,\cdots, 2i_N}\\
=&\tilde{F}^{(1)}_{1i_1}\otimes \tilde{F}^{(2)}_{1i_2}\otimes\cdots \otimes \tilde{F}^{(N)}_{1i_N}-\tilde{F}^{(1)}_{2i_1}\otimes \tilde{F}^{(2)}_{2i_2}\otimes\cdots \otimes \tilde{F}^{(N)}_{2i_N}\\
=&(\tilde{F}^{(1)}_{1i_1}-\tilde{F}^{(1)}_{2i_1})\otimes \tilde{F}^{(2)}_{1i_2}\otimes\cdots \otimes \tilde{F}^{(N)}_{1i_N}+\tilde{F}^{(1)}_{2i_1}\otimes[\tilde{F}^{(2)}_{1i_2}\otimes\cdots \otimes \tilde{F}^{(N)}_{1i_N}-\tilde{F}^{(2)}_{2i_2}\otimes\cdots \otimes \tilde{F}^{(N)}_{2i_N}],\\
\endaligned
\end{eqnarray}
by induction it is then easy to get that
\begin{eqnarray}
\aligned
&\tilde{F}_{1i_1,1i_2,\cdots, 1i_N}-\tilde{F}_{2i_1,2i_2,\cdots, 2i_N}\\
=&(\tilde{F}^{(1)}_{1i_1}-\tilde{F}^{(1)}_{2i_1})\otimes \tilde{F}^{(2)}_{1i_2}\otimes\cdots \otimes \tilde{F}^{(N)}_{1i_N}\\
&+\tilde{F}^{(1)}_{2i_1}\otimes (\tilde{F}^{(2)}_{1i_2}-\tilde{F}^{(2)}_{2i_2})\otimes \tilde{F}^{(3)}_{1i_3}\otimes \cdots \otimes \tilde{F}^{(N)}_{1i_N}\\
&+\tilde{F}^{(1)}_{2i_1}\otimes \tilde{F}^{(2)}_{2i_2}\otimes (\tilde{F}^{(3)}_{1i_3}-\tilde{F}^{(3)}_{2i_3})\otimes \cdots \otimes \tilde{F}^{(N)}_{1i_N}\\
&\vdots\\
&+\tilde{F}^{(1)}_{2i_1}\otimes \tilde{F}^{(2)}_{2i_2}\otimes \cdots \otimes (\tilde{F}^{(N)}_{1i_N}-\tilde{F}^{(N)}_{2i_N}).
\endaligned
\end{eqnarray}
Thus
\begin{eqnarray}
\aligned
&\sum_{i_1,i_2,\cdots, i_N}(\tilde{F}_{1i_1,1i_2,\cdots, 1i_N}-\tilde{F}_{2i_1,2i_2,\cdots, 2i_N})^\dagger(\tilde{F}_{1i_1,1i_2,\cdots, 1i_N}-\tilde{F}_{2i_1,2i_2,\cdots, 2i_N})\\
=&\sum_{l=1}^N I\otimes I\otimes \cdots \otimes [\sum_{i_l}(\tilde{F}^{(l)}_{1i_l}-\tilde{F}^{(l)}_{2i_l})^\dagger (\tilde{F}^{(l)}_{1i_l}-\tilde{F}^{(l)}_{2i_l})]\otimes I\otimes \cdots \otimes I\\
&+\sum_{l_2=1}^N\sum_{l_1=1}^{l_2-1}[I\otimes \cdots \otimes [\sum_{i_{l_1}}(\tilde{F}^{(l_1)}_{1i_{l_1}}-\tilde{F}^{(l_1)}_{2i_{l_1}})^\dagger \tilde{F}^{(l_1)}_{2i_{l_1}}]\otimes \cdots \otimes  [\sum_{i_{l_2}}\tilde{F}^{(l_2)\dagger}_{1i_{l_2}}(\tilde{F}^{(l_2)}_{1i_{l_2}}-\tilde{F}^{(l_2)}_{2i_{l_2}})]\otimes \cdots \otimes I+h.c]\\
=&\sum_{l=1}^N I\otimes I\otimes \cdots \otimes (2I-K_W-K_W^\dagger)\otimes I\otimes \cdots \otimes I\\
&+\sum_{l_2=1}^N\sum_{l_1=1}^{l_2-1}[I\otimes \cdots \otimes (K_W-I)\otimes \cdots \otimes  (I-K_W^\dagger)\otimes \cdots \otimes I+h.c].\\
\endaligned
\end{eqnarray}
here again $W$ is the first $q\times q$ block of $U^\dagger V$ and $K_{W}=\sum_{i=1}^q\sum_{j=1}^qw_{ij}F_{1i}^\dagger F_{2j}$ with $w_{ij}$ as the $ij$-th entry of $W$. We then have
\begin{eqnarray}
\label{eq:suppN}
\aligned
&2-2\cos \Theta_{QC}(K_1^{\otimes N},K_2^{\otimes N})\\
&\leq \|\sum_{i_1,i_2,\cdots, i_N}(\tilde{F}_{1i_1,1i_2,\cdots, 1i_N}-\tilde{F}_{2i_1,2i_2,\cdots, 2i_N})^\dagger(\tilde{F}_{1i_1,1i_2,\cdots, 1i_N}-\tilde{F}_{2i_1,2i_2,\cdots, 2i_N})\|\\
&\leq N\|2I-K_W-K_W^\dagger\|+N(N-1)\|I-K_W\|^2.
\endaligned
\end{eqnarray}
\end{widetext}


\begin{thebibliography}{41}
\expandafter\ifx\csname natexlab\endcsname\relax\def\natexlab#1{#1}\fi
\expandafter\ifx\csname bibnamefont\endcsname\relax
  \def\bibnamefont#1{#1}\fi
\expandafter\ifx\csname bibfnamefont\endcsname\relax
  \def\bibfnamefont#1{#1}\fi
\expandafter\ifx\csname citenamefont\endcsname\relax
  \def\citenamefont#1{#1}\fi
\expandafter\ifx\csname url\endcsname\relax
  \def\url#1{\texttt{#1}}\fi
\expandafter\ifx\csname urlprefix\endcsname\relax\def\urlprefix{URL }\fi
\providecommand{\bibinfo}[2]{#2}
\providecommand{\eprint}[2][]{\url{#2}}

\bibitem[{\citenamefont{Fuchs and Caves}(1994)}]{Fuchs1994}
\bibinfo{author}{\bibfnamefont{C.~A.} \bibnamefont{Fuchs}} \bibnamefont{and}
  \bibinfo{author}{\bibfnamefont{C.~M.} \bibnamefont{Caves}},
  \bibinfo{journal}{Phys. Rev. Lett.} \textbf{\bibinfo{volume}{73}},
  \bibinfo{pages}{3047} (\bibinfo{year}{1994}),
  \urlprefix\url{http://link.aps.org/doi/10.1103/PhysRevLett.73.3047}.

\bibitem[{\citenamefont{Jozsa}(1994)}]{Jozsa1994}
\bibinfo{author}{\bibfnamefont{R.}~\bibnamefont{Jozsa}},
  \bibinfo{journal}{Journal of Modern Optics} \textbf{\bibinfo{volume}{41}},
  \bibinfo{pages}{2315} (\bibinfo{year}{1994}),
  \eprint{http://dx.doi.org/10.1080/09500349414552171},
  \urlprefix\url{http://dx.doi.org/10.1080/09500349414552171}.

\bibitem[{\citenamefont{Fuchs}(1996)}]{Fuchs1996}
\bibinfo{author}{\bibfnamefont{C.~A.} \bibnamefont{Fuchs}},
  \bibinfo{journal}{arXiv} pp. \bibinfo{pages}{quant--ph/9601020}
  (\bibinfo{year}{1996}).

\bibitem[{\citenamefont{Braunstein and Caves}(1994)}]{BRAU94}
\bibinfo{author}{\bibfnamefont{S.~L.} \bibnamefont{Braunstein}}
  \bibnamefont{and} \bibinfo{author}{\bibfnamefont{C.~M.} \bibnamefont{Caves}},
  \bibinfo{journal}{Phys. Rev. Lett.} \textbf{\bibinfo{volume}{72}},
  \bibinfo{pages}{3439} (\bibinfo{year}{1994}),
  \urlprefix\url{http://link.aps.org/doi/10.1103/PhysRevLett.72.3439}.

\bibitem[{\citenamefont{Schumacher}(1996)}]{Schumacher1996}
\bibinfo{author}{\bibfnamefont{B.}~\bibnamefont{Schumacher}},
  \bibinfo{journal}{Phys. Rev. A} \textbf{\bibinfo{volume}{54}},
  \bibinfo{pages}{2614} (\bibinfo{year}{1996}),
  \urlprefix\url{http://link.aps.org/doi/10.1103/PhysRevA.54.2614}.

\bibitem[{\citenamefont{Surmacz et~al.}(2006)\citenamefont{Surmacz, Nunn,
  Waldermann, Wang, Walmsley, and Jaksch}}]{Surmacz2006}
\bibinfo{author}{\bibfnamefont{K.}~\bibnamefont{Surmacz}},
  \bibinfo{author}{\bibfnamefont{J.}~\bibnamefont{Nunn}},
  \bibinfo{author}{\bibfnamefont{F.~C.} \bibnamefont{Waldermann}},
  \bibinfo{author}{\bibfnamefont{Z.}~\bibnamefont{Wang}},
  \bibinfo{author}{\bibfnamefont{I.~A.} \bibnamefont{Walmsley}},
  \bibnamefont{and} \bibinfo{author}{\bibfnamefont{D.}~\bibnamefont{Jaksch}},
  \bibinfo{journal}{Phys. Rev. A} \textbf{\bibinfo{volume}{74}},
  \bibinfo{pages}{050302} (\bibinfo{year}{2006}),
  \urlprefix\url{http://link.aps.org/doi/10.1103/PhysRevA.74.050302}.

\bibitem[{\citenamefont{Gu}(2010)}]{Gu2010}
\bibinfo{author}{\bibfnamefont{S.-J.} \bibnamefont{Gu}}, \bibinfo{journal}{Int.
  J. Mod. Phys. B} \textbf{\bibinfo{volume}{24}}, \bibinfo{pages}{4371}
  (\bibinfo{year}{2010}).

\bibitem[{\citenamefont{Barnum et~al.}(2000)\citenamefont{Barnum, Knill, and
  Nielsen}}]{Barnum1998}
\bibinfo{author}{\bibfnamefont{H.}~\bibnamefont{Barnum}},
  \bibinfo{author}{\bibfnamefont{E.}~\bibnamefont{Knill}}, \bibnamefont{and}
  \bibinfo{author}{\bibfnamefont{M.~A.} \bibnamefont{Nielsen}},
  \bibinfo{journal}{IEEE Trans. Info. Theor.} \textbf{\bibinfo{volume}{46}},
  \bibinfo{pages}{1317} (\bibinfo{year}{2000}), \eprint{quant-ph/9809010}.

\bibitem[{\citenamefont{Kitaev}(1997)}]{Kitaev1997}
\bibinfo{author}{\bibfnamefont{A.}~\bibnamefont{Kitaev}},
  \bibinfo{journal}{Russian Mathematical Surveys}
  \textbf{\bibinfo{volume}{52}}, \bibinfo{pages}{1191} (\bibinfo{year}{1997}),
  \urlprefix\url{http://www.turpion.org/php/paper.phtml?journal_id=rm&paper_id=2155}.

\bibitem[{\citenamefont{Kitaev et~al.}(2002)\citenamefont{Kitaev, Shen, and
  Vyalyi}}]{Kitaev2002}
\bibinfo{author}{\bibfnamefont{A.~Y.} \bibnamefont{Kitaev}},
  \bibinfo{author}{\bibfnamefont{A.~H.} \bibnamefont{Shen}}, \bibnamefont{and}
  \bibinfo{author}{\bibfnamefont{M.~N.} \bibnamefont{Vyalyi}},
  \emph{\bibinfo{title}{Classical and Quantum Computation}}
  (\bibinfo{publisher}{American Mathematical Society},
  \bibinfo{address}{Boston, MA, USA}, \bibinfo{year}{2002}), ISBN
  \bibinfo{isbn}{0821832298}.

\bibitem[{\citenamefont{Watrous}(2009)}]{watrous2009}
\bibinfo{author}{\bibfnamefont{J.}~\bibnamefont{Watrous}},
  \bibinfo{journal}{arXiv} p. \bibinfo{pages}{0901.4709}
  (\bibinfo{year}{2009}).

\bibitem[{\citenamefont{Gilchrist et~al.}(2005)\citenamefont{Gilchrist,
  Langford, and Nielsen}}]{Gilchrist}
\bibinfo{author}{\bibfnamefont{A.}~\bibnamefont{Gilchrist}},
  \bibinfo{author}{\bibfnamefont{N.~K.} \bibnamefont{Langford}},
  \bibnamefont{and} \bibinfo{author}{\bibfnamefont{M.~A.}
  \bibnamefont{Nielsen}}, \bibinfo{journal}{Phys. Rev. A}
  \textbf{\bibinfo{volume}{71}}, \bibinfo{pages}{062310}
  (\bibinfo{year}{2005}),
  \urlprefix\url{http://link.aps.org/doi/10.1103/PhysRevA.71.062310}.

\bibitem[{\citenamefont{Belavkin et~al.}(2005)\citenamefont{Belavkin,
  D’Ariano, and Raginsky}}]{Belavkin}
\bibinfo{author}{\bibfnamefont{V.~P.}~\bibnamefont{Belavkin}},
  \bibinfo{author}{\bibfnamefont{G.~M.} \bibnamefont{D’Ariano}},
  \bibnamefont{and} \bibinfo{author}{\bibfnamefont{M.}
  \bibnamefont{Raginsky}}, \bibinfo{journal}{J. Mathematical Physics}
  \textbf{\bibinfo{volume}{46}}, \bibinfo{pages}{062106}
  (\bibinfo{year}{2005}).
  
\bibitem[{\citenamefont{Chau}(2011)}]{Chau2011}
\bibinfo{author}{\bibfnamefont{H.~F.} \bibnamefont{Chau}},
  \bibinfo{journal}{Quant. Inf. Comput.} \textbf{\bibinfo{volume}{11}},
  \bibinfo{pages}{0721} (\bibinfo{year}{2011}).

\bibitem[{\citenamefont{Fung and Chau}(2013)}]{Fung1}
\bibinfo{author}{\bibfnamefont{C.-H.~F.} \bibnamefont{Fung}} \bibnamefont{and}
  \bibinfo{author}{\bibfnamefont{H.~F.} \bibnamefont{Chau}},
  \bibinfo{journal}{Phys. Rev. A} \textbf{\bibinfo{volume}{88}},
  \bibinfo{pages}{012307} (\bibinfo{year}{2013}),
  \urlprefix\url{http://link.aps.org/doi/10.1103/PhysRevA.88.012307}.

\bibitem[{\citenamefont{Fung and Chau}(2014)}]{Fung2}
\bibinfo{author}{\bibfnamefont{C.-H.~F.} \bibnamefont{Fung}} \bibnamefont{and}
  \bibinfo{author}{\bibfnamefont{H.~F.} \bibnamefont{Chau}},
  \bibinfo{journal}{Phys. Rev. A} \textbf{\bibinfo{volume}{90}},
  \bibinfo{pages}{022333} (\bibinfo{year}{2014}),
  \urlprefix\url{http://link.aps.org/doi/10.1103/PhysRevA.90.022333}.

\bibitem[{\citenamefont{Acin}(2001)}]{Acin01}
\bibinfo{author}{\bibfnamefont{A.}~\bibnamefont{Acin}}, \bibinfo{journal}{Phys.
  Rev. Lett.} \textbf{\bibinfo{volume}{87}}, \bibinfo{pages}{177901}
  (\bibinfo{year}{2001}),
  \urlprefix\url{http://link.aps.org/doi/10.1103/PhysRevLett.87.177901}.

\bibitem[{\citenamefont{Yuan and Fung}(2017)}]{Yuan2017npj}
\bibinfo{author}{\bibfnamefont{H.}~\bibnamefont{Yuan}} \bibnamefont{and}
  \bibinfo{author}{\bibfnamefont{C.-H.~F.} \bibnamefont{Fung}},
  \bibinfo{journal}{to appear in npj Quantum Information}
  (\bibinfo{year}{2017}).

\bibitem{Fuchs1999}
 C. A. Fuchs and J. van de Graaf, 
{\em IEEE Trans. Inf. Theory} 45, 1216 (1999).

\bibitem{Raginsky}
Maxim Raginsky,
{\em Physics Letters A}, {\bf 290}, 11-18 (2001).

\bibitem[{\citenamefont{D'Ariano et~al.}(2001)\citenamefont{D'Ariano,
  Lo~Presti, and Paris}}]{Mauro2001}
\bibinfo{author}{\bibfnamefont{G.~M.} \bibnamefont{D'Ariano}},
  \bibinfo{author}{\bibfnamefont{P.}~\bibnamefont{Lo~Presti}},
  \bibnamefont{and} \bibinfo{author}{\bibfnamefont{M.~G.~A.}
  \bibnamefont{Paris}}, \bibinfo{journal}{Phys. Rev. Lett.}
  \textbf{\bibinfo{volume}{87}}, \bibinfo{pages}{270404}
  (\bibinfo{year}{2001}),
  \urlprefix\url{http://link.aps.org/doi/10.1103/PhysRevLett.87.270404}.

\bibitem[{\citenamefont{Uhlmann}(1976)}]{Uhlmann1976}
\bibinfo{author}{\bibfnamefont{A.}~\bibnamefont{Uhlmann}},
  \bibinfo{journal}{Rep. Math. Phys.} \textbf{\bibinfo{volume}{9}},
  \bibinfo{pages}{273-279} (\bibinfo{year}{1976}).

\bibitem[{\citenamefont{Yuan and Fung}(2015)}]{Yuan2015}
\bibinfo{author}{\bibfnamefont{H.}~\bibnamefont{Yuan}} \bibnamefont{and}
  \bibinfo{author}{\bibfnamefont{C.-H.~F.} \bibnamefont{Fung}},
  \bibinfo{journal}{arXiv} p. \bibinfo{pages}{1506.01909}
  (\bibinfo{year}{2015}).

\bibitem[{\citenamefont{Helstrom}(1976)}]{HELS67}
\bibinfo{author}{\bibfnamefont{C.~W.} \bibnamefont{Helstrom}},
  \emph{\bibinfo{title}{Quantum Detection and Estimation Theory}}
  (\bibinfo{publisher}{Academic Press}, \bibinfo{year}{1976}).

\bibitem[{\citenamefont{Holevo}(1982)}]{HOLE82}
\bibinfo{author}{\bibfnamefont{A.~S.} \bibnamefont{Holevo}},
  \emph{\bibinfo{title}{Probabilistic and Statistical Aspect of Quantum
  Theory}} (\bibinfo{publisher}{North-Holland}, \bibinfo{year}{1982}).

\bibitem[{\citenamefont{Cramer}(1946)}]{CRAM46}
\bibinfo{author}{\bibfnamefont{H.}~\bibnamefont{Cramer}},
  \emph{\bibinfo{title}{Mathematical Methods of Statistics}}
  (\bibinfo{publisher}{Princeton University,}, \bibinfo{year}{1946}).

\bibitem[{\citenamefont{Rao}(1945)}]{Rao}
\bibinfo{author}{\bibfnamefont{C.~R.~B.} \bibnamefont{Rao}},
  \bibinfo{journal}{Calcutta Math. Soc.} \textbf{\bibinfo{volume}{37}},
  \bibinfo{pages}{81} (\bibinfo{year}{1945}).

\bibitem[{\citenamefont{Braunstein et~al.}(1996)\citenamefont{Braunstein,
  Caves, and Milburn}}]{BRAU96}
\bibinfo{author}{\bibfnamefont{S.~L.} \bibnamefont{Braunstein}},
  \bibinfo{author}{\bibfnamefont{C.~M.} \bibnamefont{Caves}}, \bibnamefont{and}
  \bibinfo{author}{\bibfnamefont{G.~J.} \bibnamefont{Milburn}},
  \bibinfo{journal}{Annals of Physics} \textbf{\bibinfo{volume}{247}},
  \bibinfo{pages}{135} (\bibinfo{year}{1996}).

\bibitem[{\citenamefont{Fujiwara and Imai}(2008)}]{Fujiwara2008}
\bibinfo{author}{\bibfnamefont{A.}~\bibnamefont{Fujiwara}} \bibnamefont{and}
  \bibinfo{author}{\bibfnamefont{H.}~\bibnamefont{Imai}},
  \bibinfo{journal}{Journal of Physics A: Mathematical and Theoretical}
  \textbf{\bibinfo{volume}{41}}, \bibinfo{pages}{255304}
  (\bibinfo{year}{2008}).

\bibitem[{\citenamefont{Escher et~al.}(2011)\citenamefont{Escher,
  de~Matos~Filho, and Davidovich}}]{Escher2011}
\bibinfo{author}{\bibfnamefont{B.~M.} \bibnamefont{Escher}},
  \bibinfo{author}{\bibfnamefont{R.~L.} \bibnamefont{de~Matos~Filho}},
  \bibnamefont{and}
  \bibinfo{author}{\bibfnamefont{L.}~\bibnamefont{Davidovich}},
  \bibinfo{journal}{Nature Physics} \textbf{\bibinfo{volume}{7}},
  \bibinfo{pages}{406} (\bibinfo{year}{2011}).

\bibitem[{\citenamefont{Tsang}(2013)}]{Tsang2013}
\bibinfo{author}{\bibfnamefont{M.}~\bibnamefont{Tsang}}, \bibinfo{journal}{New
  J. Phys.} \textbf{\bibinfo{volume}{15}}, \bibinfo{pages}{073005}
  (\bibinfo{year}{2013}).

\bibitem[{\citenamefont{Demkowicz-Dobrzanski
  et~al.}(2012)\citenamefont{Demkowicz-Dobrzanski, Kolodynski, and
  Guta}}]{Rafal2012}
\bibinfo{author}{\bibfnamefont{R.}~\bibnamefont{Demkowicz-Dobrzanski}},
  \bibinfo{author}{\bibfnamefont{J.}~\bibnamefont{Kolodynski}},
  \bibnamefont{and} \bibinfo{author}{\bibfnamefont{M.}~\bibnamefont{Guta}},
  \bibinfo{journal}{Nature Communications} \textbf{\bibinfo{volume}{3}},
  \bibinfo{pages}{1063} (\bibinfo{year}{2012}).

\bibitem[{\citenamefont{Knysh et~al.}(2011)\citenamefont{Knysh, Smelyanskiy,
  and Durkin}}]{durkin}
\bibinfo{author}{\bibfnamefont{S.}~\bibnamefont{Knysh}},
  \bibinfo{author}{\bibfnamefont{V.~N.} \bibnamefont{Smelyanskiy}},
  \bibnamefont{and} \bibinfo{author}{\bibfnamefont{G.~A.}
  \bibnamefont{Durkin}}, \bibinfo{journal}{Phys. Rev. A}
  \textbf{\bibinfo{volume}{83}}, \bibinfo{pages}{021804}
  (\bibinfo{year}{2011}),
  \urlprefix\url{http://link.aps.org/doi/10.1103/PhysRevA.83.021804}.

\bibitem[{\citenamefont{Knysh et~al.}(2014)\citenamefont{Knysh, Chen, and
  Durkin}}]{Knysh2014}
\bibinfo{author}{\bibfnamefont{S.}~\bibnamefont{Knysh}},
  \bibinfo{author}{\bibfnamefont{E.}~\bibnamefont{Chen}}, \bibnamefont{and}
  \bibinfo{author}{\bibfnamefont{G.}~\bibnamefont{Durkin}},
  \bibinfo{journal}{arXiv} p. \bibinfo{pages}{1402.0495}
  (\bibinfo{year}{2014}), \urlprefix\url{http://arxiv.org/abs/1402.0495}.

\bibitem[{\citenamefont{Kolodynski and Demkowicz-Dobrzanski}(2013)}]{Jan2013}
\bibinfo{author}{\bibfnamefont{J.}~\bibnamefont{Kolodynski}} \bibnamefont{and}
  \bibinfo{author}{\bibfnamefont{R.}~\bibnamefont{Demkowicz-Dobrzanski}},
  \bibinfo{journal}{New J. Phys.} \textbf{\bibinfo{volume}{15}},
  \bibinfo{pages}{073043} (\bibinfo{year}{2013}).

\bibitem[{\citenamefont{Demkowicz-Dobrza\ifmmode~\acute{n}\else \'{n}\fi{}ski
  and Maccone}(2014)}]{Rafal2014}
\bibinfo{author}{\bibfnamefont{R.}~\bibnamefont{Demkowicz-Dobrza\ifmmode~\acute{n}\else
  \'{n}\fi{}ski}} \bibnamefont{and}
  \bibinfo{author}{\bibfnamefont{L.}~\bibnamefont{Maccone}},
  \bibinfo{journal}{Phys. Rev. Lett.} \textbf{\bibinfo{volume}{113}},
  \bibinfo{pages}{250801} (\bibinfo{year}{2014}),
  \urlprefix\url{http://link.aps.org/doi/10.1103/PhysRevLett.113.250801}.

\bibitem[{\citenamefont{Alipour et~al.}(2014)\citenamefont{Alipour, Mehboudi,
  and Rezakhani}}]{Alipour2014}
\bibinfo{author}{\bibfnamefont{S.}~\bibnamefont{Alipour}},
  \bibinfo{author}{\bibfnamefont{M.}~\bibnamefont{Mehboudi}}, \bibnamefont{and}
  \bibinfo{author}{\bibfnamefont{A.~T.} \bibnamefont{Rezakhani}},
  \bibinfo{journal}{Phys. Rev. Lett.} \textbf{\bibinfo{volume}{112}},
  \bibinfo{pages}{120405} (\bibinfo{year}{2014}),
  \urlprefix\url{http://link.aps.org/doi/10.1103/PhysRevLett.112.120405}.


\bibitem[{\citenamefont{Duan et~al.}(2007)\citenamefont{Duan, Feng, and
  Ying}}]{Duan2007}
\bibinfo{author}{\bibfnamefont{R.}~\bibnamefont{Duan}},
  \bibinfo{author}{\bibfnamefont{Y.}~\bibnamefont{Feng}}, \bibnamefont{and}
  \bibinfo{author}{\bibfnamefont{M.}~\bibnamefont{Ying}},
  \bibinfo{journal}{Phys. Rev. Lett.} \textbf{\bibinfo{volume}{98}},
  \bibinfo{pages}{100503} (\bibinfo{year}{2007}),
  \urlprefix\url{http://link.aps.org/doi/10.1103/PhysRevLett.98.100503}.

\bibitem[{\citenamefont{Duan et~al.}(2008)\citenamefont{Duan, Feng, and
  Ying}}]{Duan2008}
\bibinfo{author}{\bibfnamefont{R.}~\bibnamefont{Duan}},
  \bibinfo{author}{\bibfnamefont{Y.}~\bibnamefont{Feng}}, \bibnamefont{and}
  \bibinfo{author}{\bibfnamefont{M.}~\bibnamefont{Ying}},
  \bibinfo{journal}{Phys. Rev. Lett.} \textbf{\bibinfo{volume}{100}},
  \bibinfo{pages}{020503} (\bibinfo{year}{2008}),
  \urlprefix\url{http://link.aps.org/doi/10.1103/PhysRevLett.100.020503}.

\bibitem[{\citenamefont{Lu et~al.}(2012)\citenamefont{Lu, Chen, and
  Duan}}]{Cheng2012}
\bibinfo{author}{\bibfnamefont{C.}~\bibnamefont{Lu}},
  \bibinfo{author}{\bibfnamefont{J.}~\bibnamefont{Chen}}, \bibnamefont{and}
  \bibinfo{author}{\bibfnamefont{R.}~\bibnamefont{Duan}},
  \bibinfo{journal}{Quantum Info. Comput.} \textbf{\bibinfo{volume}{12}},
  \bibinfo{pages}{138} (\bibinfo{year}{2012}), ISSN \bibinfo{issn}{1533-7146},
  \urlprefix\url{http://dl.acm.org/citation.cfm?id=2231036.2231045}.

\bibitem[{\citenamefont{Chiribella et~al.}(2008)\citenamefont{Chiribella,
  D'Ariano, and Perinotti}}]{ChiribellaDP08}
\bibinfo{author}{\bibfnamefont{G.}~\bibnamefont{Chiribella}},
  \bibinfo{author}{\bibfnamefont{G.~M.} \bibnamefont{D'Ariano}},
  \bibnamefont{and}
  \bibinfo{author}{\bibfnamefont{P.}~\bibnamefont{Perinotti}},
  \bibinfo{journal}{Phys. Rev. Lett.} \textbf{\bibinfo{volume}{101}},
  \bibinfo{pages}{180501} (\bibinfo{year}{2008}),
  \urlprefix\url{http://link.aps.org/doi/10.1103/PhysRevLett.101.180501}.

\bibitem[{\citenamefont{Duan et~al.}(2009)\citenamefont{Duan, Feng, and
  Ying}}]{DuanFY09}
\bibinfo{author}{\bibfnamefont{R.}~\bibnamefont{Duan}},
  \bibinfo{author}{\bibfnamefont{Y.}~\bibnamefont{Feng}}, \bibnamefont{and}
  \bibinfo{author}{\bibfnamefont{M.}~\bibnamefont{Ying}},
  \bibinfo{journal}{Phys. Rev. Lett.} \textbf{\bibinfo{volume}{103}},
  \bibinfo{pages}{210501} (\bibinfo{year}{2009}),
  \urlprefix\url{http://link.aps.org/doi/10.1103/PhysRevLett.103.210501}.

\bibitem[{\citenamefont{Harrow et~al.}(2010)\citenamefont{Harrow, Hassidim,
  Leung, and Watrous}}]{Harrow2010}
\bibinfo{author}{\bibfnamefont{A.~W.} \bibnamefont{Harrow}},
  \bibinfo{author}{\bibfnamefont{A.}~\bibnamefont{Hassidim}},
  \bibinfo{author}{\bibfnamefont{D.~W.} \bibnamefont{Leung}}, \bibnamefont{and}
  \bibinfo{author}{\bibfnamefont{J.}~\bibnamefont{Watrous}},
  \bibinfo{journal}{Phys. Rev. A} \textbf{\bibinfo{volume}{81}},
  \bibinfo{pages}{032339} (\bibinfo{year}{2010}),
  \urlprefix\url{http://link.aps.org/doi/10.1103/PhysRevA.81.032339}.

\bibitem[{\citenamefont{Grant and Boyd}(2011)}]{CVX}
\bibinfo{author}{\bibfnamefont{M.}~\bibnamefont{Grant}} \bibnamefont{and}
  \bibinfo{author}{\bibfnamefont{S.}~\bibnamefont{Boyd}}
  (\bibinfo{year}{2011}), \urlprefix\url{http://cvxr.com/cvx/}.


\bibitem{Choi}
M.D. Choi,  and  C.K. Li, {\em J. Operator Theory.} {\bf 46}, 435 (2001).

\bibitem{Fung3}
C.-H. F. Fung, H. F. Chau, C.K. Li,  and N.S. Sze,
Quantum Information and Computation 15, 0685-0693 (2015).

\bibitem{ChildsPR00}
A.~Childs, J.~Preskill, and J.~Renes.
\newblock Quantum information and precision measurement.
\newblock {\em Journal of Modern Optics}, 47(2--3):155--176, 2000.

\bibitem{YuanUniversal}
H. Yuan, C.-H. F. Fung,
{\em Phys. Rev. Lett.} 115, 110401 (2015).
  
  
\end{thebibliography}
\end{document}